\documentclass[12pt]{oxarticle}
\pdfoutput=1
\usepackage{amsmath, amssymb, amsthm, mathrsfs, mathtools, amscd}
\usepackage{graphicx, longtable, float}
\usepackage{xspace}
\usepackage[latin1]{inputenc}
\usepackage[usenames,dvipsnames]{color}
\usepackage{rotating}
\usepackage{tikz}
\usetikzlibrary{arrows,shapes}

\renewcommand{\a}{\alpha}

\newcommand{\D}{\Delta}

\renewcommand{\o}{\omega}

\newcommand{\IP}{\mathbb{P}}

\newcommand{\IZ}{\mathbb{Z}}

\font\twentyonerm=cmr12 at 21pt
\font\eightrm=cmr8 at 8pt
\font\csc=cmcsc10

\newcommand{\beq}{\begin{equation}}
\newcommand{\eeq}{\end{equation}}
\newcommand{\beqnn}{\begin{equation*}}
\newcommand{\eeqnn}{\end{equation*}}

\newcommand{\fref}[1]{Figure~\ref{#1}}
\newcommand{\tref}[1]{Table~\ref{#1}}
\newcommand{\sref}[1]{Section~\ref{#1}}         

\newcommand{\cy}{Calabi--Yau\xspace}
\newcommand{\hodgenos}{(h^{1,1},\,h^{2,1})}
\newcommand{\cicy}[2]{\begin{matrix} #1\end{matrix}\!\left[\begin{matrix}#2 \end{matrix}\right]}
\newcommand{\one}{{\hskip-0.75pt\bf 1}}                                     

\def\place#1#2#3{\vbox to0pt{\kern-\parskip\kern-7pt
                             \kern-#2truein\hbox{\kern#1truein #3}
                             \vss}\nointerlineskip}
                        
\newcommand{\capt}[3]{\parbox{#1}{\renewcommand{\baselinestretch}{1.0}
                                                           \caption{\label{#2}\small\it #3}}}

\newcommand{\twoheadlongrightarrow}{\relbar\joinrel\twoheadrightarrow}

\newcommand{\Bigcheck}{\lower2pt\hbox{\smash{\hbox{{\twentyonerm \v{}}}}}}
\newcommand{\Bighat}{\lower3.8pt\hbox{\smash{\hbox{{\twentyonerm \^{}}}}}}

\newcommand{\Xsharp}{\mathscr{X}^{\raisebox{2pt}{$\scriptstyle\sharp$}}}

\newcommand{\Xcheck}{\kern2pt\hbox{\Bigcheck\kern-15pt{$\mathscr{X}$}}}
\newcommand{\Xhat}{\kern2pt\hbox{\Bighat\kern-12pt{$\mathscr{X}$}}}

\renewcommand{\baselinestretch}{1.1}

\numberwithin{equation}{section}
\proofmodefalse

\begin{document}
\pagestyle{empty}
\begin{center}
\null\vskip0.6in
{\LARGE Completing the Web of $\IZ_3$ - Quotients\\[8pt]
of Complete Intersection Calabi-Yau Manifolds\\[0.6in]}
{\csc Philip Candelas$^1$ and Andrei Constantin$^2$ \\[1in]}
{\it $^1$Mathematical Institute\hphantom{$^1$}\\
Oxford University\\
24-29 St.\ Giles'\\
Oxford OX1 3LB, UK\\[0.5in]
$^2$Rudolf Peierls Centre for Theoretical Physics\hphantom{$^2$}\\
Oxford University\\
1 Keble Road\\
Oxford OX1 4NP, UK\\}
\vfill
{\bf Abstract\\[3ex]}
\parbox{6.0in}{\setlength{\baselineskip}{14pt}
We complete the study \cite{CD08} of smooth $\IZ_3$-quotients of complete intersection Calabi-Yau threefolds by discussing the six new manifolds that admit free $\IZ_3$ actions that were discovered in \cite{Volker}. These manifolds were missed in \cite{CD08} and complete the web of smooth $\IZ_3$-quotients in a nice way. We discuss the transitions between these manifolds and include also the other manifolds of the web. This leads to the conclusion that the web of $\IZ_3$-free quotients of complete intersection Calabi-Yau threefolds is connected by conifold transitions. 
}
\end{center}

\newpage
\tableofcontents
 
\newpage
\setcounter{page}{1}
\pagestyle{plain}
\section{Introduction and Generalities}		


Non-simply connected Calabi-Yau threefolds have played an important role in the compactification of the heterotic string theory for a long time \cite{CHSW, TY}. Most of the known examples of such manifolds are quotients of complete intersection Calabi-Yau (CICY) manifolds by freely acting discrete symmetries. 

The interest in smooth quotients of CICY manifolds was renewed with the observation, made in \cite{Triadophilia}, that there is an interesting corner in the string landscape, corresponding to Calabi-Yau threefolds with small Hodge numbers. Subsequently, this corner was populated with some 80 new manifolds \cite{CD08}, constructed either as free or resolved quotients of CICY manifolds. The observation was made also that quotients with isomorphic fundamental groups form webs connected by conifold transitions. The search for CICY manifolds admitting free linear group actions was completed, for the configurations of the CICY list, by Braun \cite{Volker} by means of an automated scan, leading to a classification of all linear actions of discrete groups on the CICY manifolds constructed in \cite{Candelas88}. Together with many new examples of free quotients this search revealed also a new manifold with Euler number $-6$, leading to a physical model with three generations of particles interacting according to the  gauge group of the Standard Model. This manifold \cite{BCD}, realized as a quotient by a group of order 12, enjoys the remarkable property of having the smallest possible Hodge numbers for a manifold for which three generations of particles is achieved via the standard embedding, namely $\hodgenos = (1,4)$. 

It would be both interesting and important to give a detailed account of all the new manifolds and symmetries uncovered by \cite{Volker}. Our aim here, however, is more modest: we give instead a detailed description of the six new $\IZ_3$ quotients appearing in the list \cite{Volker}, which were missed in \cite{CD08}. This more modest goal is nevertheless worthwhile since the new $\IZ_3$ manifolds fit into the web of $\IZ_3$ manifolds in an interesting way, as may be seen by referring to \fref{Z3Web}, which shows this web, with the six new quotient manifolds indicated by red dots. 
\begin{figure}[ht]
\begin{center}
\includegraphics[width=6.5in]{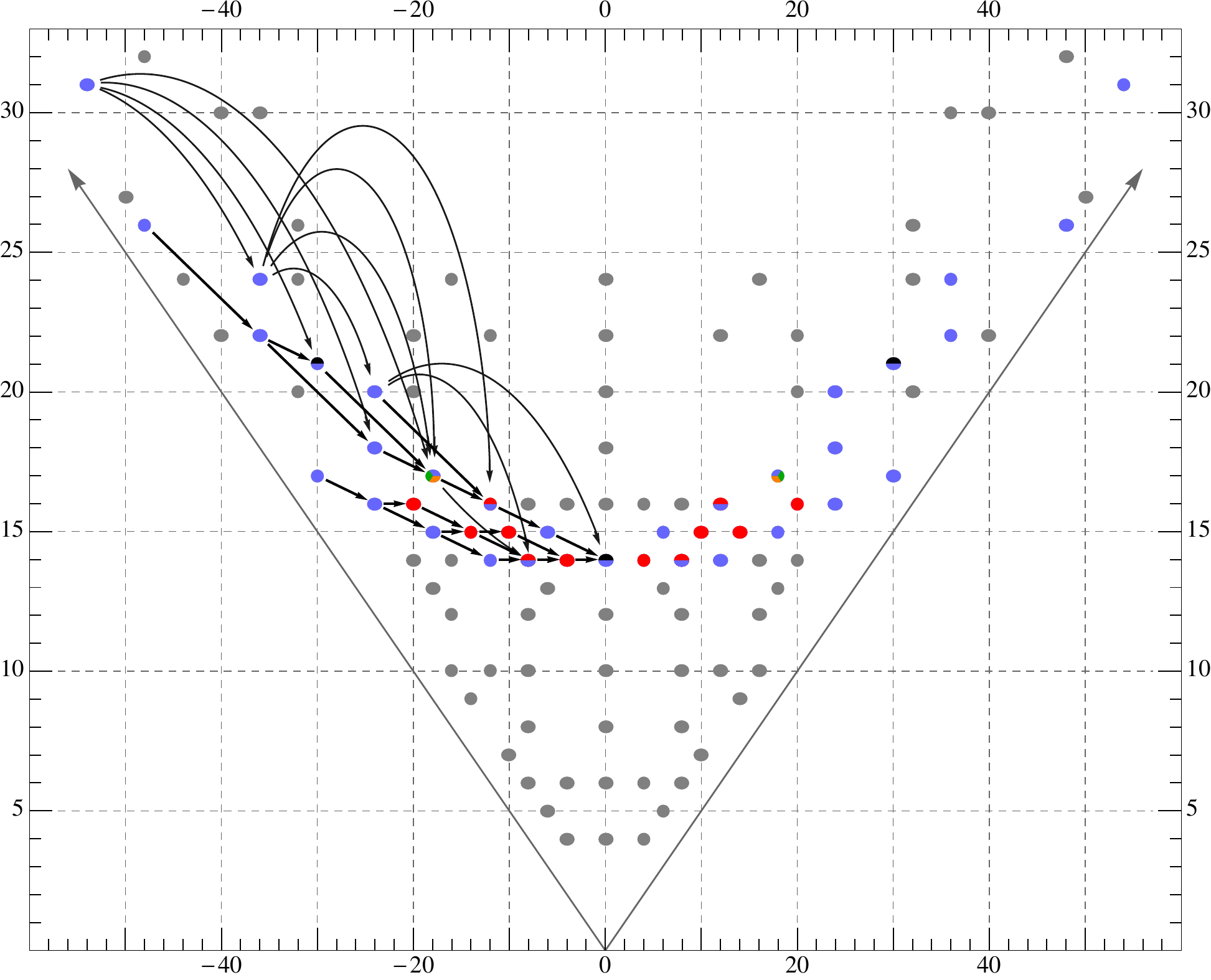}
\vskip12pt
\framebox[5.0in]{\parbox{5.0in}{\vspace{7pt}
\hspace*{20pt}\includegraphics[width=7pt]{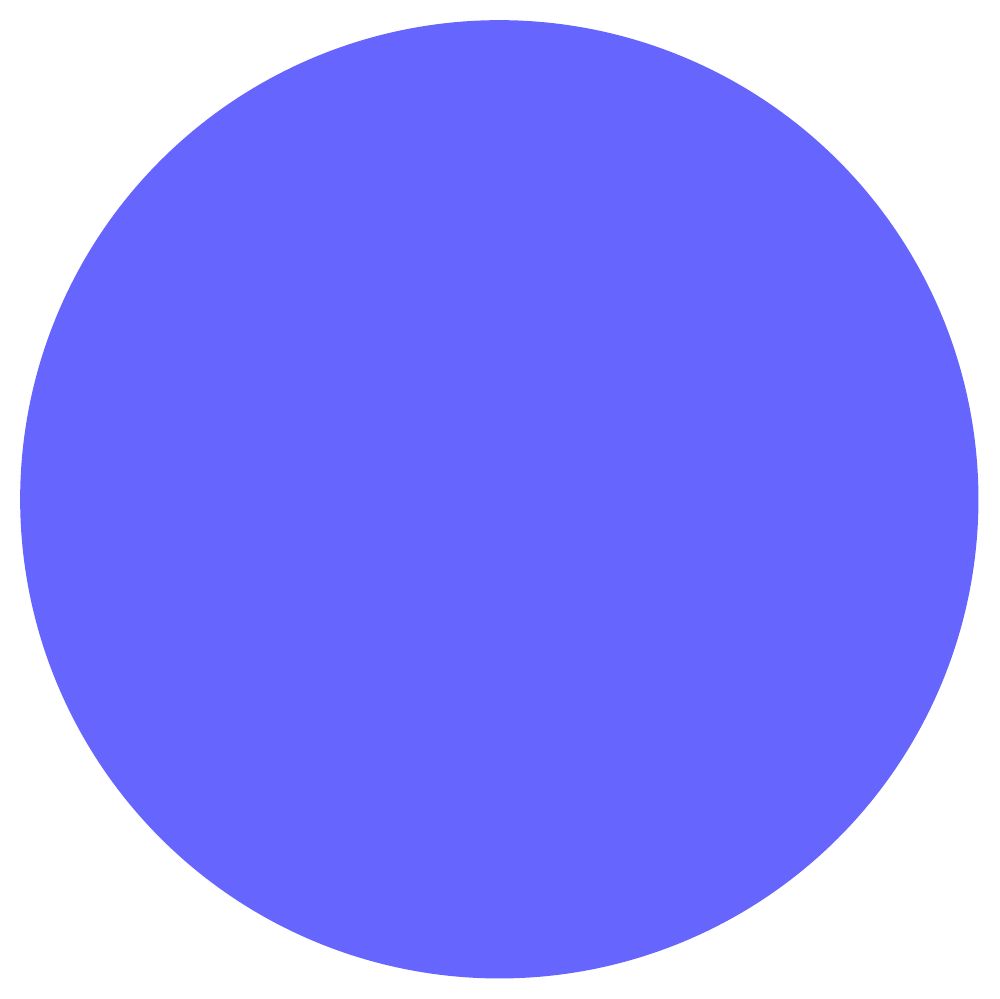}~~
\includegraphics[width=7pt]{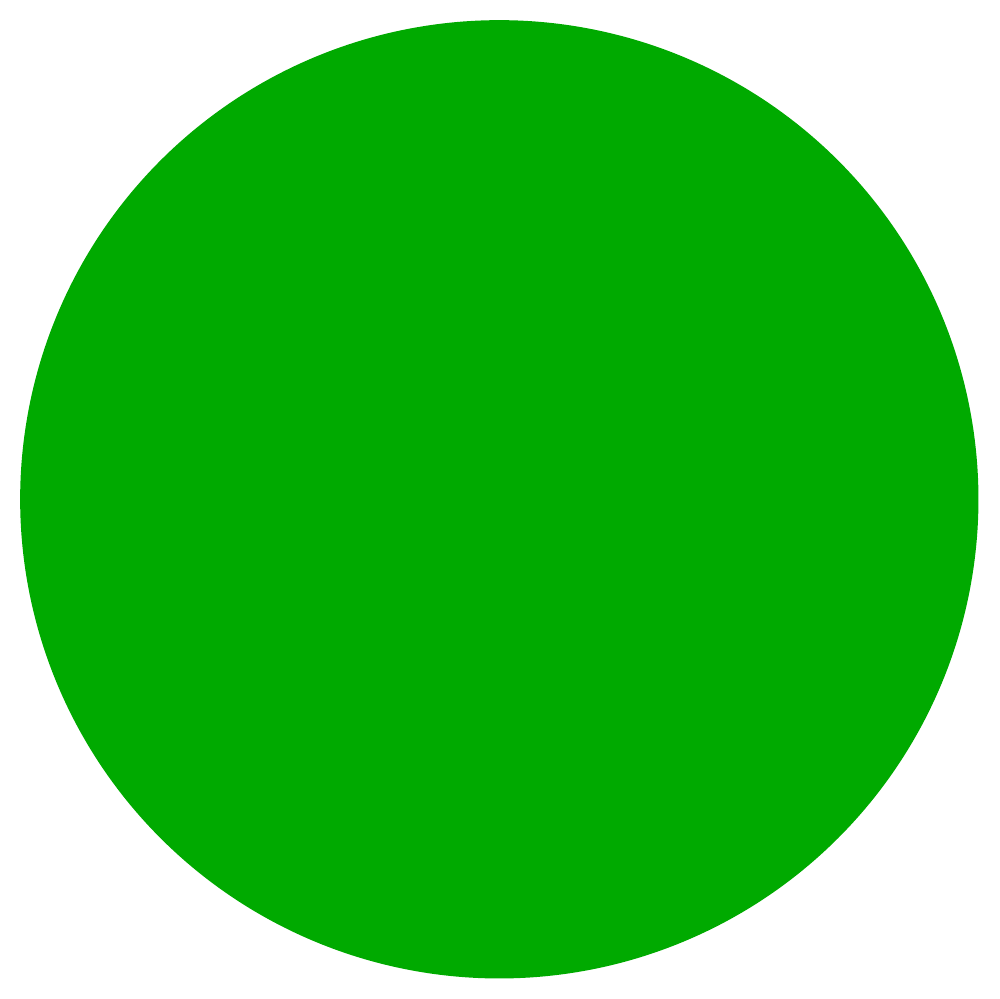}~~
\includegraphics[width=7pt]{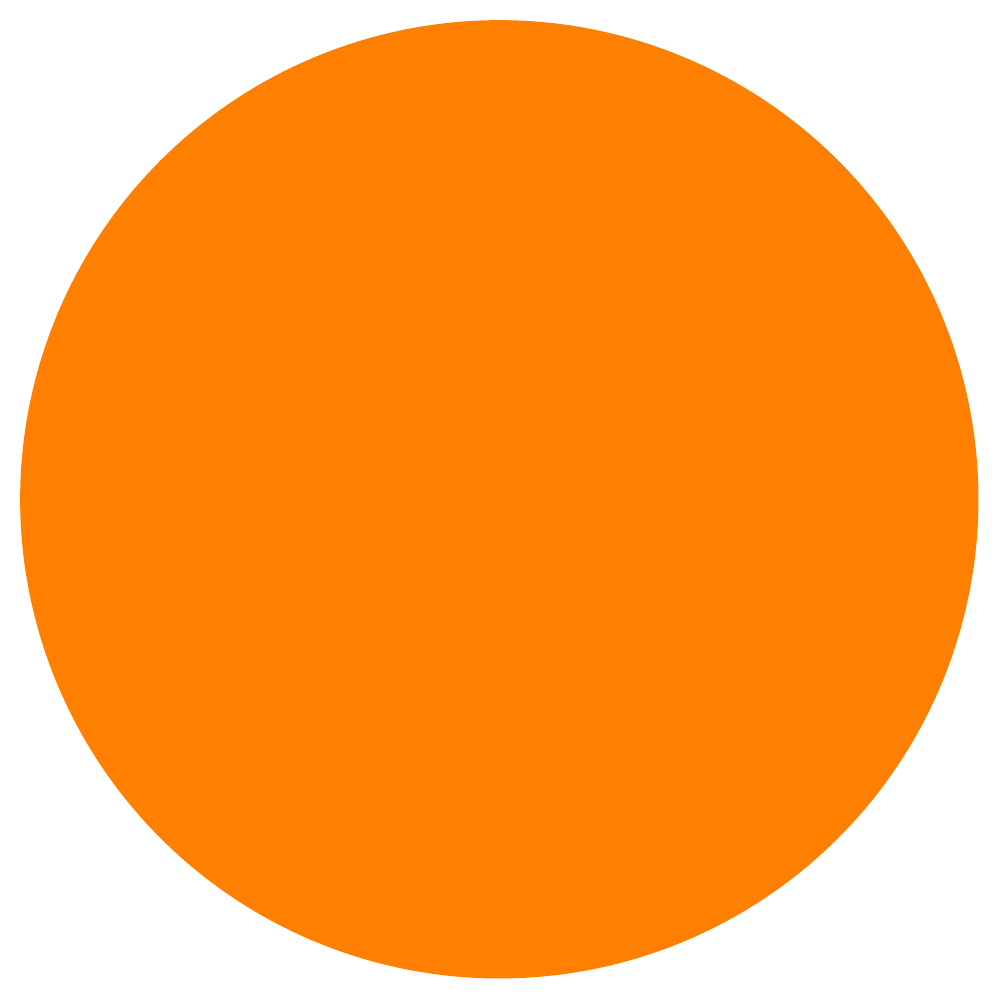}~~
\includegraphics[width=7pt]{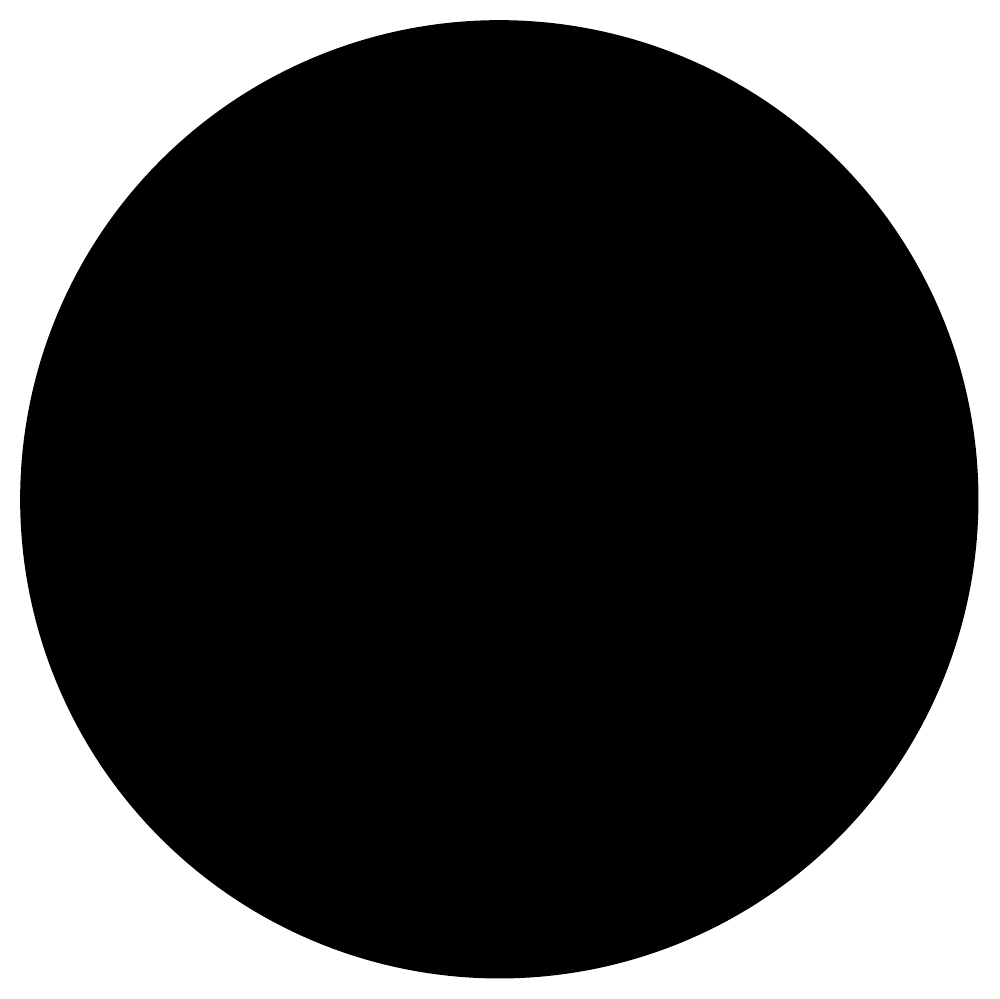}~~%
Old $\IZ_3$-free quotients of  CICY manifolds \cite{CD08, Triadophilia}. \\ 
\hspace*{76pt}\includegraphics[width=7pt]{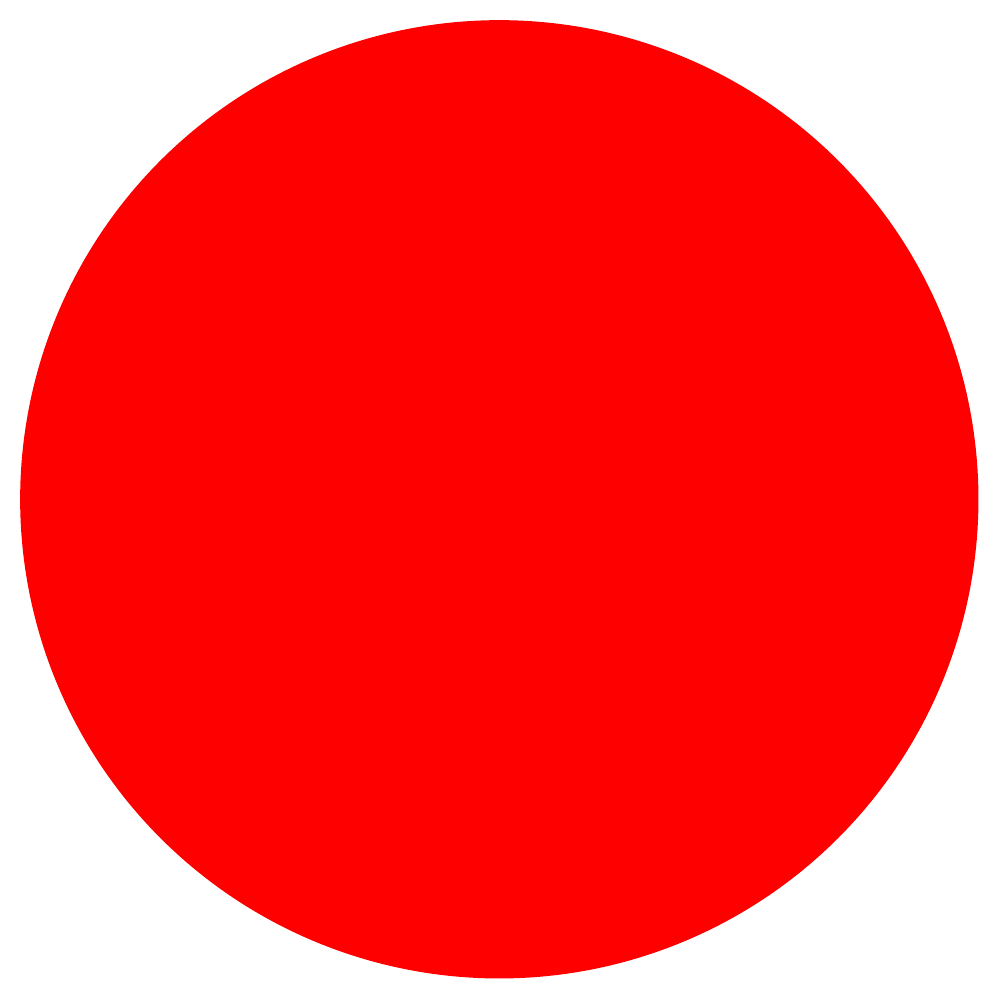}~~New $\IZ_3$-free quotients of CICY manifolds from \cite{Volker}.\\
\hspace*{34.5pt}
\includegraphics[width=7pt]{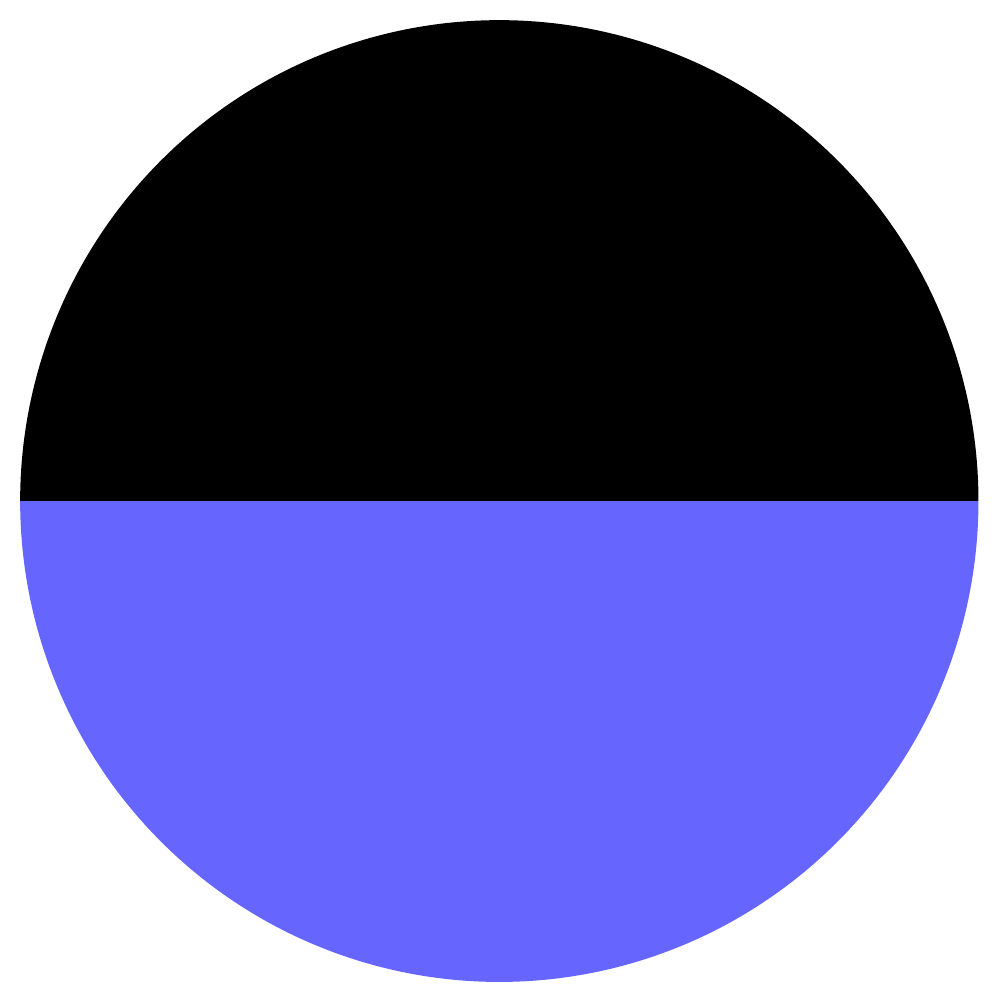}~~
\includegraphics[width=7pt]{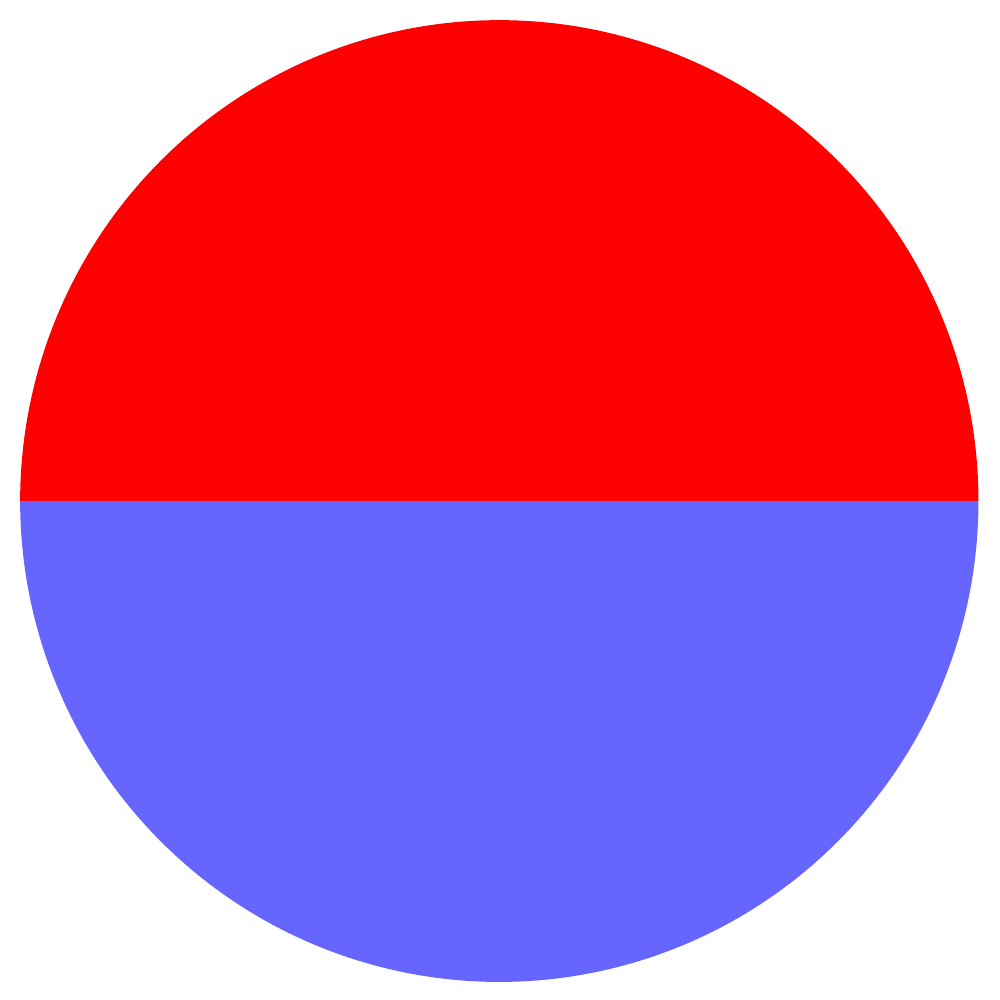}~~
\includegraphics[width=7pt]{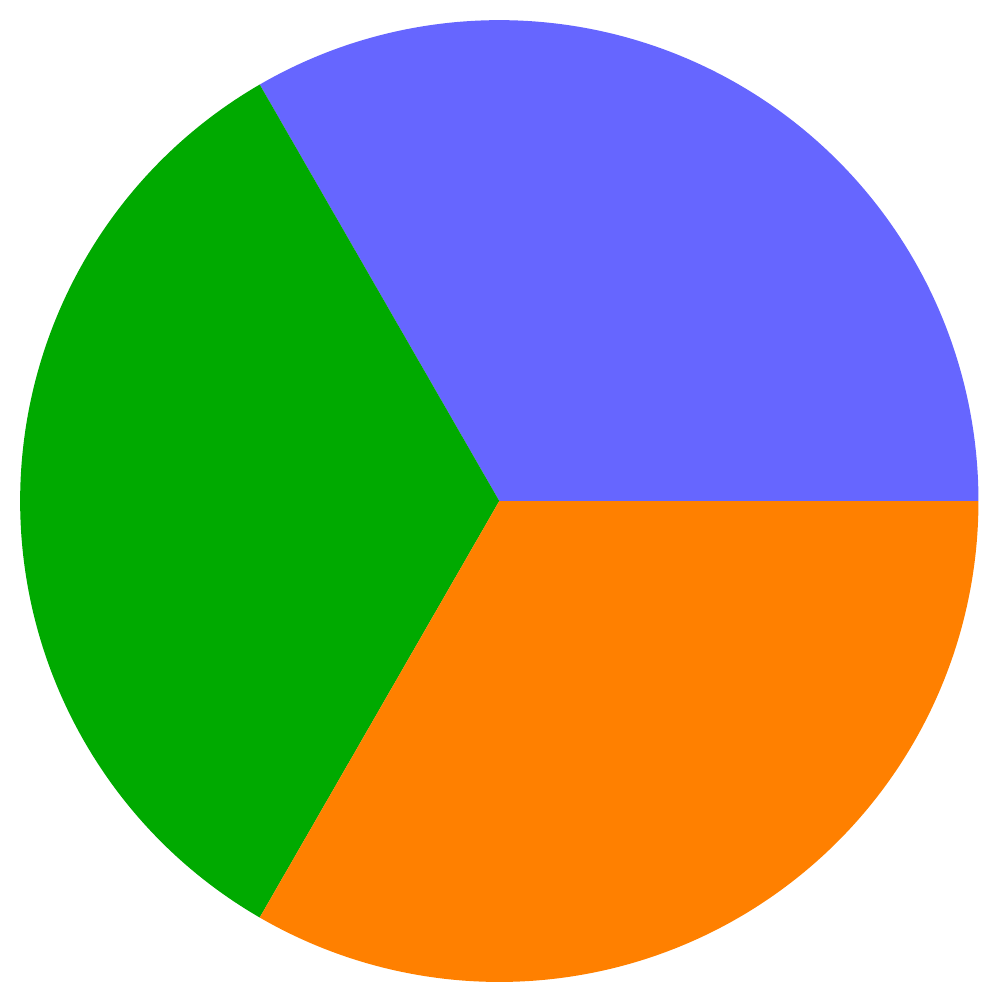}~~Multiply-occupied sites.\\[-4pt]
}}
\vskip3pt
\capt{5.5in}{Z3Web}{The web of $\IZ_3$ quotients of CICY manifolds. On the horizontal axis: the Euler number $\chi = 2\left( h^{1,1} - h^{2,1} \right)$.  On the vertical axis: the height $h^{1,1}+ h^{2,1}$.}
\end{center}
\end{figure}

In this paper we present the symmetries of the manifolds in an explicit and straightforward manner. While the group actions are given unambiguously in the results of \cite{Volker} the form in which these are given are those found by the computer programme and, while equivalent, take a very different form from that which one would naturally write. We also compute the Hodge numbers for each quotient, these are not given by the results of \cite{Volker}. On computing these, certain regularities become apparent, as we see from 
\fref{Z3Web}. In particular five of the new quotients are seen to fit, for example, into the double sequence of
\tref{doublesequenceIntro} for which the horizontal arrows correspond to $\D(h^{11},h^{21})=(1,-1)$ and the vertical arrows correspond to $\D(h^{11},h^{21})=(1,-2)$. This double sequence corresponds to the structure, evident in \fref{Z3Web}, whose top left member has coordinates $(-30,17)$. Other sequences will appear later. We have drawn \fref{Z3Web} so as to emphasise three sequences, which largely correspond to the straight arrows. These sequences are drawn with heavier lines though all the arrows represent conifold transitions.

In addition to computing the Hodge numbers we study, in \sref{secWeb}, the conifold transitions between the covering manifolds and also the conifold transitions between the quotients. We find, as in~\cite{CD08}, that the $\IZ_3$-quotients, including the new quotients, are connected by conifold transitions so as to form a single web, as we see from \fref{Z3Web}.

Our conventions follow those of \cite{CD08} and use the techniques discussed in Section 1 of that paper. For convenience, we summarize below the most important aspects concerning the construction of smooth CICY quotients (see also \cite{Hubsch}). 
\begin{table}[ht]
\begin{center}
\framebox[6.0in]{
\begin{tabular}{c}
\begin{tikzpicture}[scale=1.2]
\clip (-2.5, -.4) rectangle (9.55,5.7);
\def\nodeshadowed[#1]#2;{\node[scale=1.1,above,#1]{#2};}

\nodeshadowed [at={(-1,0 )},yslant=0.0]
{ {\small \textcolor{black} {$\mathbf{\left( \mathscr{X}^{6,24}/ \IZ_3 \right) ^{4,10}}$}} };
\nodeshadowed [at={(2,0 )},yslant=0.0]
{ {\small \textcolor{black} {$\mathbf{\color{red} \left( \mathscr{X}^{9,21}/ \IZ_3 \right) ^{5,9}}$ }} };
\nodeshadowed [at={(5,0 )},yslant=0.0]
{ {\small \textcolor{black} {$\mathbf{\color{red} \left( \mathscr{X}^{12,18}/ \IZ_3 \right) ^{6,8}}$ }} };
\nodeshadowed [at={(8,0 )},yslant=0.0]
{ {\small \textcolor{black} {$\mathbf{\left( \mathscr{X}^{15,15}/ \IZ_3 \right) ^{7,7}}$}} };
\nodeshadowed [at={(-1,1.5 )},yslant=0.0]
{ {\small \textcolor{black} {$\mathbf{\left( \mathscr{X}^{5,32}/ \IZ_3 \right) ^{3,12}}$ }} };
\nodeshadowed [at={(2,1.5 )},yslant=0.0]
{ {\small \textcolor{black} {$\mathbf{\color{red}  \left( \mathscr{X}^{8,29}/ \IZ_3 \right) ^{4,11}}$}} };
\nodeshadowed [at={(5,1.5 )},yslant=0.0]
{ {\small \textcolor{black} {$\mathbf{\color{red} \left( \mathscr{X}^{11,26}/ \IZ_3 \right) ^{5,10}}$}} };
\nodeshadowed [at={(-1,3 )},yslant=0.0]
{ {\small \textcolor{black} {$\mathbf{\left( \mathscr{X}^{4,40}/ \IZ_3 \right) ^{2,14}}$}} };
\nodeshadowed [at={(2,3 )},yslant=0.0]
{ {\small \textcolor{black} {$\mathbf{\color{red} \left(  \mathscr{X}^{7,37} / \IZ_3 \right) ^{3,13}}$}} };
\nodeshadowed [at={(-1,4.5 )},yslant=0.0]
{ {\small \textcolor{black} {$\mathbf{\left( \mathscr{X}^{3,48}/ \IZ_3 \right) ^{1,16}}$}} };

\draw[very thick,blue,->] (-1, 4.5) -- (-1,3.7);
\draw[very thick,blue,->] (-1, 3.) -- (-1,2.2);
\draw[very thick,blue,->] (-1, 1.5) -- (-1,.7);
\draw[very thick,blue,->] (2, 3.) -- (2,2.2);
\draw[very thick,blue,->] (2, 1.5) -- (2,.7);
\draw[very thick,blue,->] (5, 1.5) -- (5,.7);

\draw[very thick,blue,->] (-.1, 3.3) -- (.8, 3.3);
\draw[very thick,blue,->] (-.1, 1.8) -- (.8, 1.8);
\draw[very thick,blue,->] (-.1, .3) -- (.8, .3);
\draw[very thick,blue,->] (2.9, 1.8) -- (3.74, 1.8);
\draw[very thick,blue,->] (2.9, .3) -- (3.74, .3);
\draw[very thick,blue,->] (5.9, .3) -- (6.8, .3);

\end{tikzpicture}
\end{tabular}}

\vskip 8pt
\framebox[6.0in]{\parbox{5.5in}{\vspace{-8pt}
\begin{center}
 For the covering spaces:   
$\Delta_{\raisebox{-3pt}{$\scriptstyle{}\hskip-2pt\color{blue}\rightarrow $}} \hodgenos =\left(3, -3\right)
\,;~\Delta_{\color{blue}\downarrow}  \hodgenos = \left( 1, -8 \right) $ \\[4pt]
For the quotient spaces: 
$\Delta_{\raisebox{-3pt}{$\scriptstyle{}\hskip-2pt\color{blue}\rightarrow $}} \hodgenos =\left(1, -1\right)
\,;~\Delta_{\color{blue}\downarrow}  \hodgenos = \left( 1, -2 \right) $ 
\end{center}
\vspace{-8pt}}}
\capt{6.0in}{doublesequenceIntro}{The first sub-web of $\IZ_3$-free CICY quotients. In red, the new quotients.}
\end{center}
\end{table}
\subsection{Important aspects of CICY quotients}
\vskip -8pt
Let $\mathscr{X}$ be a Calabi-Yau manifold and $G \times {\mathscr X} \rightarrow {\mathscr X}$ an action of the finite group $G$ on ${\mathscr X}$. If $G$ acts freely and holomorphically, ${\mathscr X} / G$ will inherit the structure of a complex manifold from $\mathscr X$. Moreover, ${\mathscr X} / G$ inherits a nowhere vanishing holomorphic $n$-form, where $n$ is the complex dimension of $\mathscr X$, and consequently it is Calabi-Yau. 

Also, note that $\mathscr X$ is a covering space for the quotient ${\mathscr X} / G$, and thus the quotient ${\mathscr X} / G$ is multiply connected, unless $G$ is trivial and $\mathscr X$ simply connected.  In particular, if $\mathscr X$ is simply connected, the fundamental group of ${\mathscr X} / G$ is isomorphic to $G$. Since CICY manifolds are simply connected, all the $\IZ_3$-quotients discussed here will have fundamental group $\IZ_3$. 

Furthermore, the order of the group $G$ divides the following indices of $\mathscr X$: the Euler characteristic, the holomorphic Euler characteristic, the Hirzebruch signature and the index of the Dirac operator. These divisibility properties follows by expressing the indices as integrals of densities over the manifold $\mathscr X$. This is an important necessesary condition for the existence of a free group action since the order of the group must divide the GCD of the four indices.

If $\mathscr X$ is a CICY manifold embedded in a product of projective spaces $\mathscr A = \IP^{n_1}\times \dots \times \IP^{n_m}$ and the holomorphic action $G \times {\mathscr X} \rightarrow {\mathscr X}$ comes from an action of $G$ on the ambient space $\mathscr A$, then $G$ must preserve the projectivity of the homogeneous coordinates: 
\begin{align*}
\left( \left[ x^0_1: \dotsc :x^{n_1}_1\right], \dots, \left[x^0_m: \dotsc : x^{n_m}_m\right]  \right)  \!\xmapsto {g\in G}  &~g\!\cdot\! \left( \left[ x^0_1: \dotsc : x^{n_1}_1\right], \dots, \left[x^0_m: 
\dotsc : x^{n_m}_m\right]  \right)\\[3pt]
 \cong  &~g\!\cdot \!\left( \left[ \lambda_1 x^0_1: \dotsc : \lambda_1 x^{n_1}_1\right], \dots, \left[\lambda_m x^0_m : \dotsc :\lambda_m x^{n_m}_m\right]  \right) 
\end{align*} 

This condition is clearly satisfied by all linear action of $G$. But in general, there may exist also nonlinear projective actions. For example, in a previous classification \cite{BD07} of quotients of the split bicubic manifold given by the configuration matrix
$$
{\mathscr X}^{19, 19}~=~~
\cicy{\IP^1 \\ \IP^2\\ \IP^2}
{ 1& 1\\ 3 & 0\\ 0& 3 \\}_{0}^{19,19}
$$ 
the largest symmetry group had order 9. In \cite{BCD}, however, it was shown that the split bicubic manifold admits free linear actions of two groups of order 12, when the manifold is represented as a complete intersection embedded in a product of seven $\IP^1$ spaces.  To our knowledge, nonlinear actions on CICY manifolds have not been studied systematically. The nonlinear actions that are currently known are all related to linear actions on equivalent CICY configurations with larger ambient spaces.

The quotients considered in the following sections will always come from linear projective actions. These are, in general, combinations of  internal actions on the coordinates of an individual projective space $\IP^n$, and permutations of the projective spaces, as they occur in~$\mathscr A$. 

The list of \cite{Volker} records all free linear actions on complete intersection Calabi-Yau manifolds. More precisely, it indicates what symmetries exist for each of the 7890 classes of CICY manifolds constructed in \cite{Candelas88}. The list of CICY manifolds is available on the Calabi-Yau home page \cite{CYHP} or at \cite{DigitalCICYList}, the latter having appended Hodge numbers and other topological indices.  

The information we take from \cite{Volker} is simply the indication that a certain subclass of the CICY deformation class $\mathscr X$ admits a $\IZ_3$ symmetry. Further, we construct this action by identifying the $\IZ_3$ invariant polynomials whose complete intersection define $\mathscr X$. We check that this action is free. Finally, we compute the Hodge numbers for the quotients using the methods of \cite{CD08}, as follows: 

The Hodge number $h^{1,1}$ of K\"ahler structure parameters for the quotient ${\mathscr X}/G$ is computed by finding a representation in which $\mathscr X$ is embedded in a product of $h^{1,1}\left( \mathscr X\right)$ projective spaces. In this case, the $h^{1,1}\left( {\mathscr X}\right)$ linearly independent forms in the cohomology group $H^{1,1}\left( \mathscr X\right)$ arise as pullbacks from the hyperplane classes of $\mathscr A$ and the action of $G$ on $H^{1,1}\left( \mathscr X\right)$ is determined by its action on the ambient space $\mathscr A$.  

The number $h^{2,1}\left( \mathscr{X}/G \right)$ of complex structure parameters will be computed by counting the independent parameters in the $G$-invariant polynomials defining $\mathscr X$. By a theorem 
from~\cite{Deformations}, the method is guaranteed to work whenever the parameter count for the covering space $\mathscr{X}$ is equal to $h^{2,1}(\mathscr{X})$  and the diagram associated with the CICY manifold 
$\mathscr X$ cannot be disconnected by cutting a single leg. 

As noted above, the quotient $\mathscr{X} /G$ is a smooth manifold if and only if $G$ acts freely on $\mathscr X$ and the manifold $\mathscr X$ is smooth. In Section 2 we will check, in each case, that the intersection of $\mathscr X$  with the fixed point set of the action $G\times \mathscr{A} \rightarrow \mathscr A$ is empty. The smoothness assumption for $\mathscr X$ is equivalent to the condition that the $G$-invariant polynomials  
$\{p_j\}_{j= 1,..,K}$ 
defining $\mathscr X$ are transverse, in other words $dp_1\wedge \dots \wedge dp_K \neq 0$ on the intersection $\left\{ p_j = 0\right\}_{j= 1,..,K}\,$. We check that the equations 
$dp_1\wedge \dots \wedge dp_K = 0$ and $\left\{ p_j = 0\right\}_{j= 1,..,K}\,$ have no common solution by performing a Groebner basis calculation, as explained in \cite{CD08}. The computation is implemented in Mathematica 7.0 and uses the computer algebra system SINGULAR \cite{Singular} by means of the STRINGVACUA package \cite{Stringvacua}.

In many situations, there exist several different embeddings of the same manifold $\mathscr X$. For the purpose of computing the number of K\"ahler structure parameters, we will consider the embedding of $\mathscr X$ in a product of $h^{1,1}\left( \mathscr X\right)$ projective spaces. However, for checking transversality, we will always prefer an embedding in a product of projective spaces with fewer factors, owing to the fact that the complexity of the Groebner basis calculation increases rapidly with the number of projective spaces.  

\subsection{Conifold transitions}
\vskip -8pt
It is known \cite{GreenHuebsch, CGH} that the parameter spaces of CICY threefolds intersect along loci corresponding to conifolds, and thus form a connected web. In particular, any configuration matrix, defining a particular smooth deformation class of CICY manifolds can be brought into the form of any other configuration through a sequence of operations known as `splittings' and `contractions'. For example, if $\cicy{\mathcal{A}\ }{\, \mathcal{M}\,; c\,}$ is the configuration matrix describing the class $\mathscr X$, the deformation class $\Xcheck$ obtained from $\mathscr X$ by `splitting' the last polynomial with a $\IP^n$ space is defined by the operation
\begin{equation*}
\mathscr{X} = \cicy{\mathcal{A}\,}{\, \mathcal{M}\,; c\,}\ \twoheadlongrightarrow\ \Xcheck ~=~ 
\cicy{\IP^n \\  \mathcal{A\ }}{\mathbf{0} & 1 & 1 & \dots & 1 \\ 
\mathcal{M} & c_1 & c_2 & \dots & c_{n+1}}
\end{equation*}
where $c$ is a column vector and $\sum_{i=1}^{n+1} c_i = c$. The reversed process is called a contraction (see, e.g.~\cite{Hubsch, CGH}). 

Geometrically, such operations correspond to conifold transitions. The number of nodes of the associated conifold 
$\Xsharp$ is given as half the difference of the Euler characteristics of $\Xsharp$ and  $\Xcheck$ \cite{CGH, Conifold}.  In \sref{secWeb} we prove that if both $\mathscr X$ and $\Xcheck$ admit $G$-free quotients, and the actions on $\mathscr X$ and $\Xcheck$ reduce to the same action on $\Xsharp$, then the conifold transition commutes with taking quotients. It is known that the set of all CICY's form a web connected by conifold transitions \cite{Hubsch}. What we have observed here is that also the set of $\IZ_3$ free quotients of CICY's is also similarly connected and that the new $\IZ_3$ quotients fit into this web.  


\newpage
\section{The new $\IZ_3$ quotients}
In this section we give a detailed description of the six new $\IZ_3$-symmetric manifolds. For each of these new manifolds we present $\IZ_3$-covariant polynomials together with the group action. We check that the polynomials are transverse and that the group action is fixed point free and, finally, we compute the Hodge numbers for the quotient manifold.
\subsection{The manifold $\mathscr{X}^{12,18}$ with quotient 
$\mathscr{X}^{6,8} = \mathscr{ X}^{12,18}/ \IZ_3$} \label{sec:1.1}
\vskip -8pt
This class of manifolds is described by the following configuration matrix
\begin{equation*}\label{X12,18}
\mathscr{X}^{12,18}~=~~
\cicy{\IP^1 \\ \IP^1\\  \vrule height10pt width0pt depth8pt  \IP^1\\ \IP^1\\ \IP^1\\ \vrule height10pt width0pt depth8pt  \IP^1\\ \IP^2\\ \IP^2\\ \vrule height10pt width0pt depth8pt  \IP^2\\ \IP^2}
{ \one& 0& 0&~ 0& 0 & 0&~0&0& 0&~ \one& 0 \\
 0& \one& 0&~ 0 & 0 & 0 &~0&0& 0&~ \one& 0 \\
\vrule height10pt width0pt depth8pt  0& 0& \one&~ 0& 0&0&~0&0& 0&~ \one& 0 \\
 0& 0& 0&~ \one& 0&0&~0&0& 0&~ 0& \one \\
 0& 0& 0&~ 0& \one& 0&~ 0&0 & 0&~ 0& \one \\
\vrule height10pt width0pt depth8pt   0& 0& 0&~ 0& 0& \one &~0&0& 0&~ 0& \one \\
\one& 0& 0&~ \one&0 &0&~\one& 0& 0&~ 0&0 \\
0& \one& 0&~ 0& \one&0&~0& \one& 0&~ 0& 0 \\
\vrule height10pt width0pt depth8pt  0& 0& \one&~ 0& 0&\one&~0& 0& \one&~ 0&0 \\
0& 0& 0&~ 0& 0&0&~\one&\one& \one&~ 0& 0 \\}_{-12}^{12,18}
\end{equation*}

corresponding to the following equivalent diagrams
\vspace{-15pt}
\begin{flushright}
$$
{\includegraphics[width=2.4in]{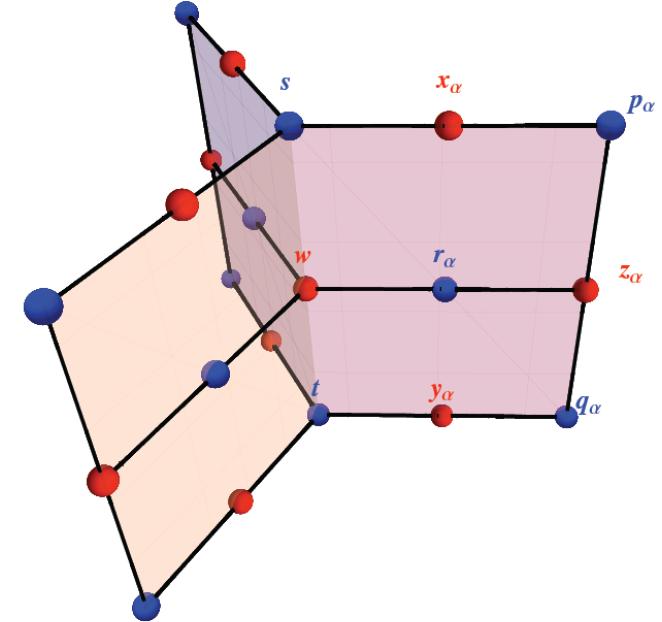}}
\hskip 32pt \lower -20pt\hbox{\includegraphics[width=2.4in]{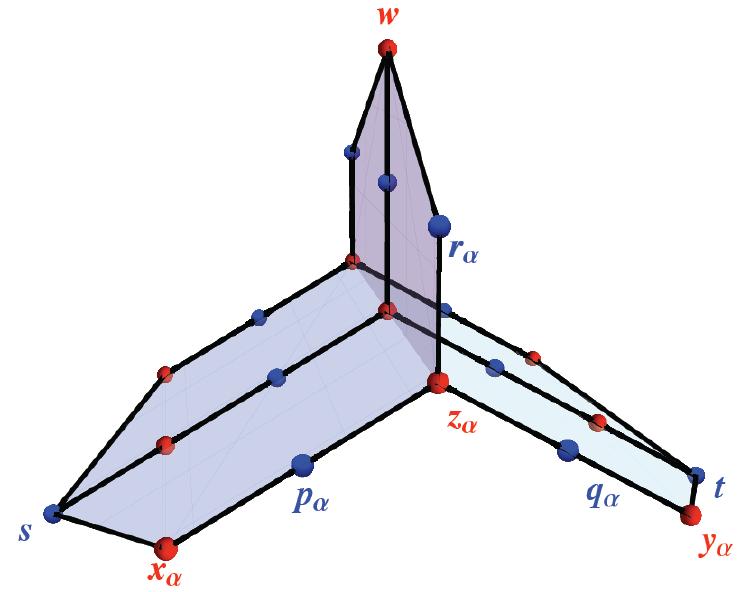}}
$$
\end{flushright}

Let us denote the coordinates of the first three $\IP^1$'s by  $x_{\alpha, j}$ and the coordinates of the remaining three $\IP^1$'s by $y_{\alpha, j}$. Here $\alpha$ labels the space and $j$ its coordinate, these labels are understood to take values in $\IZ_3$ and $\IZ_2$ respectively. Let $z_{\alpha, a}$ denote the coordinates of the first three $\IP^2$'s and $w_a$ the coordinates of the last $\IP^2$, where now the indices $\alpha$ and $a$ both take values in $\IZ_3$. 

Let us also denote by $ p^{\alpha}$ the first three polynomials, and by $q^{\alpha}$ the next three. Denote also by $s$ and $t$ the next two cubic polynomials in $\IP^1$ coordinates, and by $r^{\alpha}$ the last three cubic polynomials in $\IP^2$ coordinates. These polynomials have the general form: 
\beqnn
\begin{split}
p^{\alpha} &= \sum_{j,a} P^\a_{ja}\, x_{\a, j}\, z_{\alpha, a}~,~~~\,~~~~
q^{\alpha} = \sum_{j,a} Q^\a_{ja}\, y_{\alpha,j}\, z_{\alpha,a}\\[5pt]
s &= \sum_{i,j,k} S_{ijk}\, x_{0,i}\, x_{1,j}\, x_{2,k}~,~~~~
t = \sum_{i,j,k} T_{ijk}\, y_{0,i}\, y_{1,j}\, y_{2,k}\\[5pt]
&~~~~~~~~~~~~~~r^{\alpha} = \sum_{a,b} R^\a_{ab}\, z_{\alpha,a}\, w_{\a+b} 
\end{split}
\eeqnn
\vskip -7pt
Counting all the terms that appear in the polynomials, there are 79 coefficients in total.  Of these, 
$18=6{\times}\left(4{-}1 \right)$ coefficients can be absorbed by coordinate redefinitions of the six $\IP^1$ spaces, $32 = 4{\times}\left(9{-}1 \right)$ by redefinitions of the coordinates of the four $\IP^2$ spaces and $11$ by rescalings of the polynomials.  

This parameter count shows that there are $79{-}61 = 18$ free parameters in the polynomials, which agrees with the dimension of the $(2,1)$--cohomology group for the manifold $\mathscr{X}^{12,18}$. Thus all deformations of the complex structure of $\mathscr{X}^{12,18}$ are parametrized by polynomials and so 
$h^{2,1}\left(\mathscr{X}^{12,18} / \IZ_3\right)$ can be found through a parameter count. 

The $\IZ_3$-action is generated by
\vspace{-20pt}
\begin{center}
\begin{minipage}{0.9\textwidth}
\begin{align*}
g:~x_{\alpha, j} & ~\to~  x_{\alpha+1, j}~,& y_{\alpha, j} &~\to~ y_{\alpha+1, j}~,& z_{\a,a} &~\to~ z_{\a +1, a} ~,& w_{a} &~\to~ w_{a+1} \\[5pt]
 p^{\alpha}&~\to~  p^{\alpha+1}~,& q^{\alpha}&~\to~  q^{\alpha+1}~,& r ^{\a}&~\to~ r^{\a+1}~, & s&~\to~ s~,~~~~~ t~\to~ t ~ . 
\end{align*}
\end{minipage}
\end{center}
and we require that the polynomials $p,\,q,\,r,\,s,\,t$ are covariant under this action. This implies that the coefficients $P^\a_{ja},\, Q^\a_{ja},\, R^\a_{ab}$ are independent of $\a$ and that $S_{ijk}$ and $T_{ijk}$ are invariant under cyclic permutations of indices. 

The action is fixed point free. In the embedding space, the fixed points of $g$ are of the form $x_{\a,j} = x_j$, $y_{\a,j} = y_j$, $z_{\a,a} = z_a$ and 
$w_a \in \{\left( 1,1,1 \right), \left(1, \o, \o^2 \right), \left( 1, \o^2, \o\right) \}$, where $\o$ is a nontrivial cube root of unity. The 5 independent polynomials $p,q,r,s,t$ become constraints on the coordinates of the product 
$\IP^1{\times} \IP^1{\times}\IP^2$ parametrized by $x_j, y_j, z_a$. As such, a fixed point is a simultaneous solution of 
$$
p(x_j, z_a) ~=~q(y_j, z_a) ~=~ r(z_a) ~=~ s(x_j) ~=~ t(y_j) ~=~ 0~.
$$  
In general, there are no such solutions. Suppose $x_j$ and $y_j$ are given. Then the constraints reduce to a system of three equations for $(z_0, z_1, z_2) \in \IP^2$ which for general coefficients have no solutions. 

The parameter count for the quotient manifold goes as follows. There are $29$ terms in the $\IZ_3$ covariant form of the polynomials. Since the $\IZ_3$ action identifies the first three $\IP^1$ spaces among themselves, and similarly for the last three $\IP^1$ spaces, the number of coordinate redefinitions associated with these spaces is reduced to $6$. Also, there are only $8$ coordinate redefinitions associated with the first three $\IP^2$ spaces and 2 associated with redefinitions in the last $\IP^2$ space, which, up to scale transformations have the form: 
$$
w_{a} \to \sum_{b}\Xi_{b}\, w_{a+b}~. 
$$
The number of scale transformations for the polynomials is now 5. Summing up, the number of free parameters that describe the $\IZ_3$-symmetric class of manifolds is $29-21 = 8$, so that
$h^{2,1}\left(\mathscr{X}^{12,18}/ \IZ_3\right)  =  8$. 
\begin{table}
\begin{center}
\boxed{\hskip2pt
\begin{tabular}{cccc}
$\ \ $& & & $\ $\\
$\ \ $&$\cicy{\IP^1\\ \IP^1\\ \vrule height10pt width0pt depth8pt  \IP^1\\ \IP^1\\ \IP^1\\  \vrule height10pt width0pt depth8pt \IP^1\\ \IP^2\\ \IP^2\\ \vrule height10pt width0pt depth8pt  \IP^2\\ \IP^2\\ \IP^2\\ \IP^2}
{\one ~ 0 ~ 0 ~~ 0~ 0 ~ 0  ~~ 0 ~ 0 ~ 0 ~~ \one ~ 0 ~ 0~ ~ 0 ~ 0 ~ 0 \\
 0 ~ \one ~ 0 ~~ 0 ~ 0 ~ 0  ~~ 0 ~ 0 ~ 0~ ~ 0 ~ \one ~ 0~ ~ 0 ~ 0 ~ 0 \\
 \vrule height10pt width0pt depth8pt  0 ~ 0 ~ \one~ ~ 0 ~ 0 ~ 0  ~~ 0 ~ 0 ~ 0 ~~ 0 ~ 0 ~ \one ~~ 0 ~ 0 ~ 0  \\
 0 ~ 0 ~ 0 ~~ \one ~ 0 ~ 0 ~ ~ 0 ~ 0 ~ 0 ~~  0 ~ 0 ~ 0 ~~ \one ~ 0 ~ 0 \\
 0 ~ 0 ~ 0 ~~ 0 ~ \one ~ 0 ~ ~ 0 ~ 0 ~ 0 ~~  0 ~ 0 ~ 0~ ~ 0 ~ \one ~ 0  \\
 \vrule height10pt width0pt depth8pt  0 ~ 0 ~ 0 ~~ 0 ~ 0 ~ \one ~ ~ 0 ~ 0 ~ 0~ ~  0 ~ 0 ~ 0~ ~ 0 ~ 0 ~ \one  \\
 
\one~ 0 ~ 0~ ~ \one ~ 0 ~ 0~ ~ \one ~ 0 ~ 0 ~~ 0 ~ 0 ~ 0 ~~ 0 ~ 0 ~ 0  \\
 0 ~ \one ~ 0~ ~ 0 ~ \one ~ 0~ ~ 0 ~ \one ~ 0~ ~ 0 ~ 0 ~ 0~ ~ 0 ~ 0 ~ 0\\
\vrule height10pt width0pt depth8pt   0 ~ 0 ~ \one~ ~ 0 ~ 0 ~  \one~ ~ 0 ~ 0 ~ \one ~~0 ~ 0 ~ 0~ ~ 0 ~ 0 ~ 0 \\
 0 ~ 0 ~ 0 ~~ 0 ~ 0 ~ 0 ~~ \one ~ \one ~ \one~ ~  0 ~ 0 ~ 0~~ 0 ~ 0 ~ 0 \\
 0 ~ 0 ~ 0 ~~ 0 ~ 0 ~ 0 ~ ~0 ~ 0 ~ 0 ~ ~ \one ~ \one ~ \one~ ~ 0 ~ 0 ~ 0\\
 0 ~ 0 ~ 0 ~~ 0 ~ 0 ~ 0 ~ ~0 ~ 0 ~ 0 ~~ 0 ~ 0 ~ 0 ~~ \one ~ \one ~ \one}_{-12}^{12,18}$
&\hskip 21pt\lower 94pt\hbox{\includegraphics[width=2.4in]{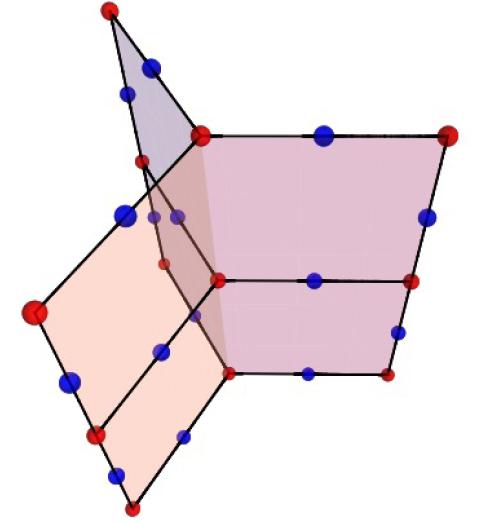}}& $\ $\\
$\ \ $& & & $\ $
\end{tabular}}
\vskip5pt
\capt{5.5in}{tab(12,18)Extended}{The matrix and diagram for the extended representation of 
$\mathscr{X}^{12,18}$ for which all 12 K\"ahler forms are represented by ambient spaces. The three vanes are identified by the $\IZ_3$ action, leaving only $6$ independent hyperplane classes.} 
\end{center}
\end{table}
The dimension of the $(1,1)$--cohomology group can be obtained by considering an extended representation of the manifold $\mathscr{X}^{12,18}$ embedded in a product of $12$ projective spaces, providing the $12$ independent $(1,1)$--forms in $H^{1,1}\left( \mathscr{X}^{12,18}\right)$. In the $\IZ_3$ quotient only $6$ of these $(1,1)$--forms remain independent (see Table \ref{tab(12,18)Extended}). 

We have checked that the polynomial constraints are transverse. Since the action of $g$ is fixed point free, we obtain a smooth quotient manifold.  The new manifold 
$\mathscr{X}^{6,8} = \mathscr{X}^{12,18}/ \IZ_3$ has fundamental group $\IZ_3$ and 
$\hodgenos = (6,8)$.
%
%
\subsection{The manifold $\mathscr{X}^{9, 21}$ with quotient 
$\mathscr{X}^{5,9} = \mathscr{X}^{9,21}/ \IZ_3$} \label{sec:1.2}
\vspace{-8pt}
The manifolds in this class are described by the configuration matrix
$$
\mathscr{X}^{9,21}~=~~
\cicy{\IP^1 \\ \IP^1\\ \vrule height10pt width0pt depth8pt  \IP^1\\ \IP^2\\ \IP^2\\ \vrule height10pt width0pt depth8pt  \IP^2\\ \IP^2\\ \IP^2}
{ \one& 0& 0& ~0& 0& 0&~0& 0& 0&~ \one \\
 0& \one& 0& ~0& 0&0&~0& 0& 0&~ \one \\
\vrule height10pt width0pt depth8pt  0& 0& \one&~ 0& 0&0&~0&0& 0&~ \one \\
\one & 0& 0&~ \one& 0&0&~\one& 0& 0&~ 0 \\
 0& \one & 0&~ 0& \one& 0&~ 0& \one& 0&~ 0 \\
 \vrule height10pt width0pt depth8pt 0& 0& \one&~ 0& 0& \one &~0&0& \one&~ 0 \\
0& 0& 0 &~ \one& \one &\one &~ 0& 0& 0&~0 \\
0& 0& 0&~ 0& 0&0&~\one& \one& \one&~ 0 \\}_{-24}^{9,21}
$$
corresponding to the equivalent diagrams
\vspace{-16pt}
\begin{flushright}
$$
{\includegraphics[width=2.4in]{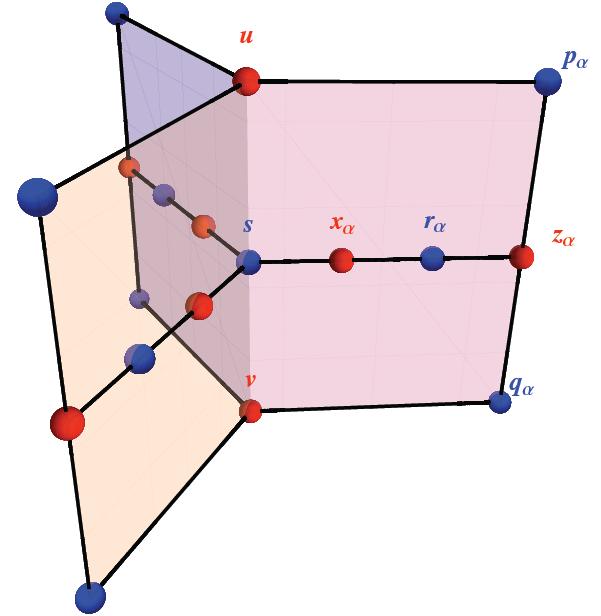}}
\hskip 32pt \lower -14pt\hbox{\includegraphics[width=2.4in]{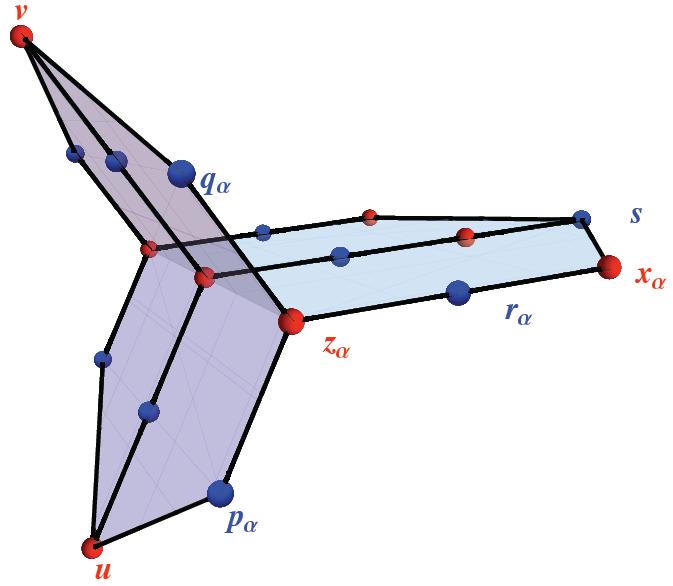}}
$$
\end{flushright}
\vspace{-5pt}

Denote the coordinates of the three $\IP^1$'s by  $x_{\alpha, j}$. As before, $\alpha$ labels the space and $j$ its coordinate, and they are understood to take values in $\IZ_3$ and $\IZ_2$ respectively. Let $z_{\alpha, a}$ denote the coordinates of the first three $\IP^2$'s, while $u_a$ and $v_a$ denote the coordinates of the last two~$\IP^2$'s.  Denote also by $ r^{\alpha}$ the first three polynomials, by $p^{\alpha}$ the next three, and  the last three by 
$q^{\a}$. Denote by $s$ the cubic polynomial in the $\IP^1$ coordinates. The polynomials can be expressed in the general form  
\vspace{-20pt}
\begin{center}
\begin{minipage}{0.7\textwidth}
\begin{align*}
p^{\alpha} ~&=~ \sum_{a,b} P^\a_{ab}\, z_{\a, a}\, u_{\alpha+b}~, & q^{\alpha} ~&=~ \sum_{a,b} Q^\a_{ab}\, z_{\alpha,a}\, v_{\alpha+b} \\[8pt]
r^{\alpha} ~&=~ \sum_{j,a} R^\a_{ja}\, x_{\alpha,j}\, z_{\a,a}~, & s ~&=~ \sum_{i,j,k} S_{ijk}\, x_{0,i}\, x_{1,j}\, x_{2,k}
\end{align*}
\end{minipage}
\end{center}
\vskip -10pt
The parameter count is similar to the previous situation. There are 80 terms in the polynomials, 49 coordinate redefinitions, and 10 scale transformations for the polynomials. This gives a total of $80-59 = 21$ free parameters in the polynomials, equal to the dimension of $H^{2,1}\left(\mathscr{X}^{9,21} \right)$. So again, we can find the value of $h^{2,1}$ for the quotient manifold by counting the free parameters of the covariant polynomials after the $\IZ_3$ identifications. 

The $\IZ_3$-action is generated by
\vspace{-25pt}
\begin{center}
\begin{minipage}{0.8\textwidth}
\begin{align*}
g:~x_{\alpha, j} & ~\to~  x_{\alpha+1, j}~,&  z_{\a,a} &~\to~ z_{\a +1, a} ~,& u_{a} &~\to~ u_{a+1}~,& v_{a} &~\to~ v_{a+1} \\[5pt]
 p^{\alpha}&~\to~  p^{\alpha+1}~,& q^{\alpha}&~\to~  q^{\alpha+1}~,& r ^{\a}&~\to~ r^{\a+1}~, & s&~\to~ s~. 
\end{align*}
\end{minipage}
\end{center}
\vspace{-5pt}
As before, in order to have $g$-covariant polynomials, we require that $P^\a_{ab},\, Q^\a_{ab},\, R^\a_{ja}$ are independent of $\a$ and that $S_{ijk}$ is invariant under cyclic permutations of indices. 

The action is fixed point free. Fixed points of $g$, in the embedding space, are of the form $x_{\a,j} = x_j$, $z_{\a,a} = z_a$ and $u_a = \xi^a$, $v_a = \zeta^a$, where $\xi$ and $\zeta$ are cube roots of unity, $\xi^3 = \zeta^3 =1$. The 4 independent polynomials $p(z_a),~ q(z_a),~ r(x_j, z_a),~ s(x_j) = 0$ become constraints on the coordinates of the product $\IP^1{\times}\IP^2$ parametrized by $x_j, z_a$. For fixed $x_j$, the constraints reduce to a system of three equations for $(z_0, z_1, z_2) \in \IP^2$ which have no solutions  for general choices of the coefficients. 

The $\IZ_3$ covariant polynomials have $28$ coefficients in total. There are also $15$ coordinate redefinitions consistent with the $\IZ_3$ identifications and $4$ scale transformations of the polynomials. This leads to 
$h^{2,1}\left(  \mathscr{X}^{9,21} / \IZ_3 \right) = 9$. 
\begin{table}[ht]
\begin{center}
\boxed{\begin{tabular}{cc}
 & \\
\lower55pt\hbox{\includegraphics[width=2.4in]{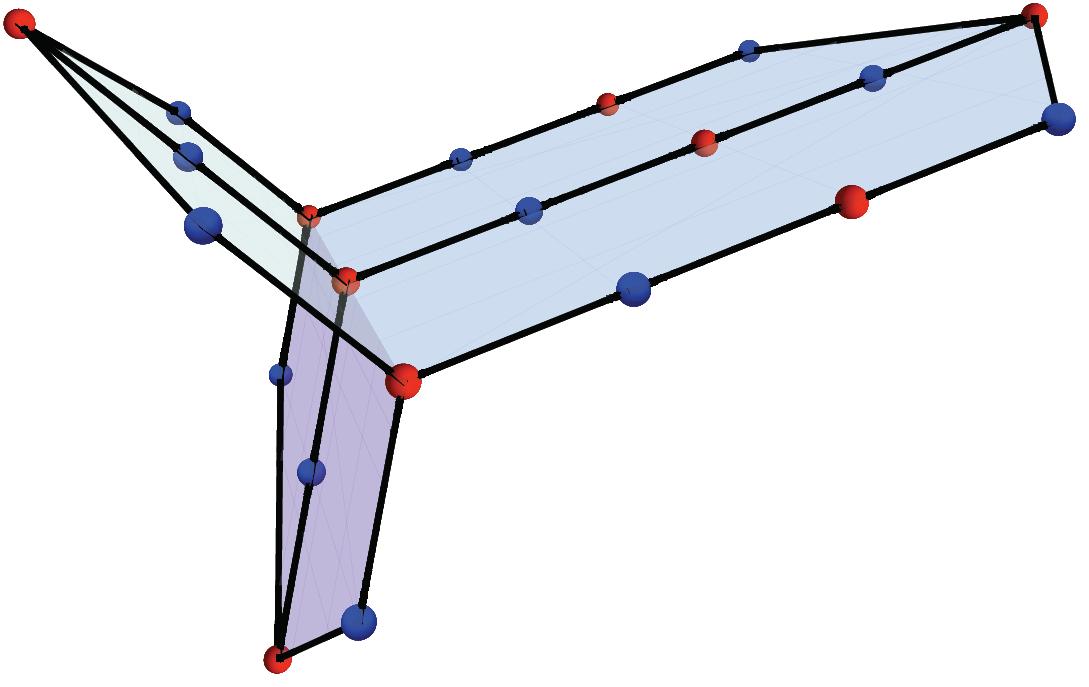}}&
\hskip .1in $\cicy{\IP^1 \\ \IP^1\\ \vrule height10pt width0pt depth8pt  \IP^1\\ \IP^2\\ \IP^2\\ \vrule height10pt width0pt depth8pt  \IP^2\\ \IP^2\\ \IP^2\\ \IP^2}
{ \one& 0& 0& ~0& 0& 0&~0& 0& 0&~ \one& 0 & 0 \\
 0& \one& 0& ~0& 0&0&~0& 0& 0&~ 0 &\one& 0 \\
\vrule height10pt width0pt depth8pt  0& 0& \one&~ 0& 0&0&~0&0& 0&~ 0 &0&\one \\
\one & 0& 0&~ \one& 0&0&~\one& 0& 0&~ 0 & 0 & 0\\
 0& \one & 0&~ 0& \one& 0&~ 0& \one& 0&~ 0 & 0 &0\\
 \vrule height10pt width0pt depth8pt 0& 0& \one&~ 0& 0& \one &~0&0& \one&~ 0 & 0 &0 \\
0& 0& 0 &~ \one& \one &\one &~ 0& 0& 0&~0 & 0 & 0 \\
0& 0& 0&~ 0& 0&0&~\one& \one& \one&~ 0 & 0 & 0\\
0& 0& 0&~ 0& 0&0&~0&0& 0&~ \one& \one & \one \\}_{-24}^{9,21}$\\
 & 
 \vspace{-4pt}
\end{tabular}}
\parbox{5.5in}{\capt{5.5in}{tab(9,21)Extended}{The matrix and diagram for the extended representation of $\mathscr{X}^{9,21}$, for which all 9 K\"ahler forms are represented by ambient spaces. After the $\IZ_3$ identification, only $5$ of the $(1,1)$--forms remain independent.}}
\end{center}
\end{table}
\vskip -14pt
The dimension of the $(1,1)$--cohomology group can be obtained by considering the extended representation of the manifold $\mathscr{X}^{9,21}$ as shown in Table \ref{tab(9,21)Extended}. It follows that $h^{1,1} = 5$ for the quotient manifold. 

We have checked that the polynomials are transverse. Thus we obtain a smooth $\IZ_3$ quotient of a CICY manifold, $\mathscr{X}^{5,9} =   \mathscr{X}^{9,21} / \IZ_3$, with fundamental group isomorphic to $\IZ_3$ and $\hodgenos = (5,9)$.
 

\subsection{The manifold $\mathscr{X}^{11, 26}$ with quotient 
$\mathscr{X}^{5,10} =  \mathscr{X}^{11,26}/ \IZ_3$} \label{sec:1.3}
\vskip -8pt
This class of manifolds is described by the configuration
$$
\mathscr{X}^{11,26}~=~~
\cicy{\IP^1 \\ \IP^1\\ \vrule height10pt width0pt depth8pt \IP^1\\ \IP^1\\ \IP^1\\\vrule height10pt width0pt depth8pt  \IP^1\\ \IP^2\\ \IP^2\\\vrule height10pt width0pt depth8pt  \IP^2}
{ \one& 0& 0&~ 0& 0 & 0&~ \one&0& 0  \\
 0& \one& 0&~ 0 & 0 & 0 &~ \one& 0& 0 \\
\vrule height10pt width0pt depth8pt  0& 0& \one&~ 0& 0&0&~ \one & 0& 0 \\
 0& 0& 0&~ \one& 0&0&~ 0 & \one& 0\\
 0& 0& 0&~ 0& \one& 0&~ 0 & \one& 0 \\
\vrule height10pt width0pt depth8pt   0& 0& 0&~ 0& 0& \one & ~ 0 & \one& 0 \\
\one& 0& 0&~ \one&0 &0&~ 0&0 & \one\\
0& \one& 0&~ 0& \one&0&~ 0& 0 & \one\\
\vrule height10pt width0pt depth8pt  0& 0& \one&~ 0& 0&\one&~ 0&0 & \one\\}_{-30}^{11,26}
$$
with equivalent diagrams
\vspace{-25pt}
\begin{flushright}
$$
{\includegraphics[width=2.4in]{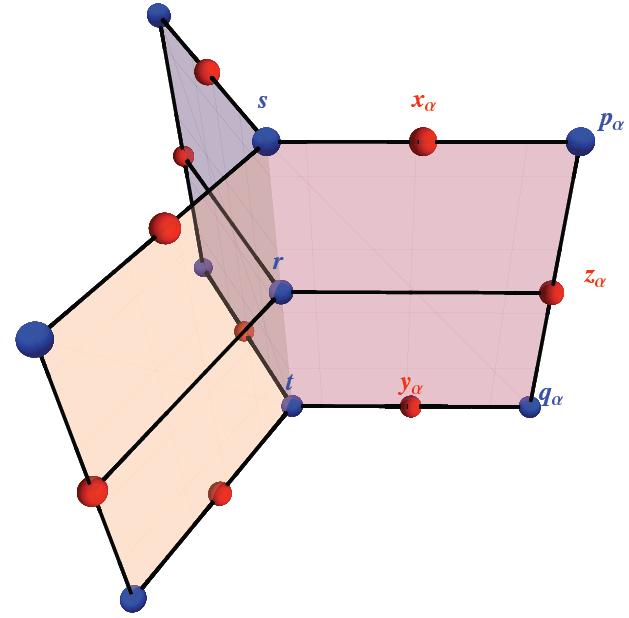}}
\hskip 32pt \lower -20pt\hbox{\includegraphics[width=2.4in]{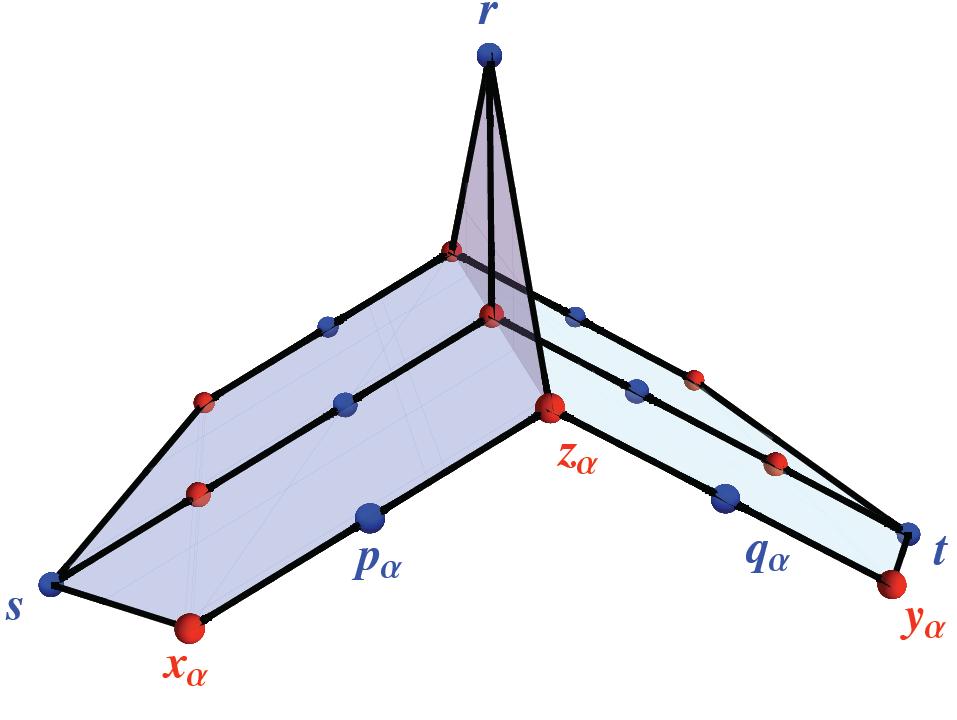}}
$$
\end{flushright}
\vspace{-7pt}

The manifold $\mathscr{X}^{11,26}$ is obtained by contracting the last $\IP^2$ in the configuration matrix of the manifold $\mathscr{X}^{12,18}$ discussed in Section \ref{sec:1.1}. In analogy with that case, we write down the following generic polynomials:
\begin{equation*}
\begin{split}
p^{\alpha} &= \sum_{j,a} P^\a_{ja}\, x_{\a, j}\, z_{\alpha, a}~,~~~~q^{\alpha}= \sum_{j,a} Q^\a_{ja}\, y_{\alpha,j}\, z_{\alpha,a}\\[8pt]
r = \sum_{a,b,c} R_{abc}\, z_{0,a}\, z_{1,b}& \, z_{2,c}~,~~~~s = \sum_{i,j,k} S_{ijk}\, x_{0,i}\, x_{1,j}\, x_{2,k}~,~~~~t = \sum_{i,j,k} T_{ijk}\, y_{0,i}\, y_{1,j}\, y_{2,k}
\end{split}
\end{equation*}

A parameter count shows that there are $79$ terms in this most general form of the polynomials. Of these, $42$ parameters can be eliminated by coordinate redefinitions, and $9$ by rescaling each of the polynomials. There are, however, two more degrees of freedom in the redefinition of $r$. These come from the following fact. For fixed $z$ coordinates, the polynomials $p^{\a} = 0$ can be regarded as a system of three equations linear in the $x$ coordinates. Using the freedom of coordinate redefinition, there are only three independent $x$ coordinates. If the equations $p^{\a} = 0$ have nontrivial solutions, we must have $\det\left.\left( \partial p / \partial x \right)\right|_z = 0$. But this is a trilinear polynomial in the $z$ coordinates and adding a multiple of it to $r$ is an allowed redefinition. Similarly, there is a second degree of freedom for $r$ associated with the vanishing of $\det\left.\left( \partial q / \partial y \right)\right|_z$. This count confirms the value of $h^{2,1} = 26 = 79 - 42 - 11$. 

The $\IZ_3$-action is generated by

\vspace{-20pt}
\begin{center}
\begin{minipage}{0.9\textwidth}
\begin{align*}
g:~~x_{\alpha, j} & ~\to~  x_{\alpha+1, j}~,& y_{\alpha, j} &~\to~ y_{\alpha+1, j}~,& z_{\a, a} &~\to~ z_{\a, a+1}\\[5pt]
 p^{\alpha}&~\to~  p^{\alpha+1}~,& q^{\alpha}&~\to~  q^{\alpha+1}~,& r &~\to~ r~, & s&~\to~ s~,~~~~~~ t~\to~ t ~ . 
\end{align*}
\end{minipage}
\end{center}

Covariance under this action requires that $P^\a_{ab},\, Q^\a_{ab},\, R^\a_{ja}$ are independent of $\a$ and that $S_{ijk}$ and $T_{ijk}$ are invariant under cyclic permutations of indices. 
The action is again fixed point~free. 

After the $\IZ_3$ identification, there are 31 possible terms in the polynomials, 14 coordinate redefinitions and 7 degrees of freedom associated with redefinitions of polynomials. These give a total of 10 free parameters. Thus 
$h^{2,1} = 10$ for the quotient manifold. 
\begin{table}[ht]
\begin{center}
\boxed{\hskip2pt
\begin{tabular}{cccc}
$\ \ $& & &$\ \ $ \\
$\ \ $&$\cicy{\IP^1\\ \IP^1\\ \vrule height10pt width0pt depth8pt  \IP^1\\ \IP^1\\ \IP^1\\  \vrule height10pt width0pt depth8pt \IP^1\\ \IP^2\\ \IP^2\\ \vrule height10pt width0pt depth8pt  \IP^2\\ \IP^2\\ \IP^2}
{\one ~ 0 ~ 0 ~~ 0~ 0 ~ 0  ~~ \one ~ 0 ~ 0~ ~ 0 ~ 0 ~ 0 ~~0\\
 0 ~ \one ~ 0 ~~ 0 ~ 0 ~ 0  ~~  0 ~ \one ~ 0~ ~ 0 ~ 0 ~ 0 ~~0\\
 \vrule height10pt width0pt depth8pt  0 ~ 0 ~ \one~ ~ 0 ~ 0 ~ 0  ~~ 0 ~ 0 ~ \one ~~ 0 ~ 0 ~ 0  ~~0\\
 0 ~ 0 ~ 0 ~~ \one ~ 0 ~ 0 ~ ~   0 ~ 0 ~ 0 ~~ \one ~ 0 ~ 0 ~~0\\
 0 ~ 0 ~ 0 ~~ 0 ~ \one ~ 0 ~ ~   0 ~ 0 ~ 0~ ~ 0 ~ \one ~ 0  ~~0\\
 \vrule height10pt width0pt depth8pt  0 ~ 0 ~ 0 ~~ 0 ~ 0 ~ \one ~ ~  0 ~ 0 ~ 0~ ~ 0 ~ 0 ~ \one ~~0 \\
 
\one~ 0 ~ 0~ ~ \one ~ 0 ~ 0 ~~ 0 ~ 0 ~ 0 ~~ 0 ~ 0 ~ 0  ~~1\\
 0 ~ \one ~ 0~ ~ 0 ~ \one ~ 0~ ~ 0 ~ 0 ~ 0~ ~ 0 ~ 0 ~ 0~~1\\
\vrule height10pt width0pt depth8pt   0 ~ 0 ~ \one~ ~ 0 ~ 0 ~  \one ~~0 ~ 0 ~ 0~ ~ 0 ~ 0 ~ 0 ~~1\\
 0 ~ 0 ~ 0 ~~ 0 ~ 0 ~ 0 ~~ \one ~ \one ~ \one~ ~  0 ~ 0 ~ 0~~ 0 \\
 0 ~ 0 ~ 0 ~~ 0 ~ 0 ~ 0 ~ ~0 ~ 0 ~ 0 ~ ~ \one ~ \one ~ \one~ ~ 0 }_{-30}^{11,26}$
&\hskip 21pt\lower 44pt\hbox{\includegraphics[width=2.4in]{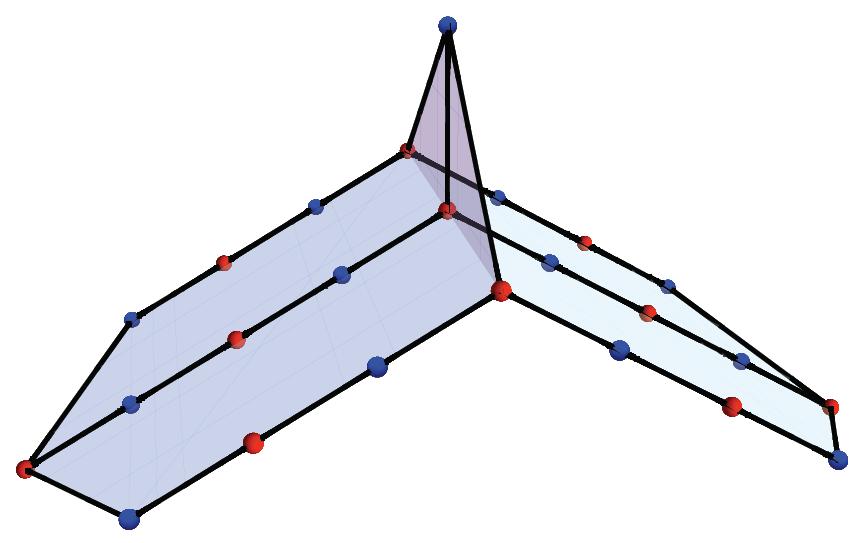}}&$\ \ $\\
$\ \ $& & &$\ \ $
\end{tabular}}
\vskip5pt
\capt{5.5in}{tab(11,26)Extended}{The matrix and diagram for the extended representation of 
$\mathscr{X}^{11,26}$ for which all 11 K\"ahler forms are represented by ambient spaces. After the $\IZ_3$ identification, only $5$ of the $(1,1)$--forms remain independent.} 
\end{center}
\end{table}
\vspace{-8pt}
In order to find $h^{1,1}$, consider the extended representation of the manifold, obtained by splitting $s$ and $t$ with two $\IP^2$'s, as shown in Table \ref{tab(11,26)Extended}. In this way we see that $h^{1,1} = 5$ for the quotient manifold. We have also checked that the polynomials are transverse. Thus, we obtain a quotient 
$\mathscr{X}^{5,10} = \mathscr{X}^{11,26}/ \IZ_3$ with fundamental group $\IZ_3$. 


\subsection{The manifold $\mathscr{X}^{8, 29}$ with quotient 
$\mathscr{X}^{4, 11} =  \mathscr{X}^{8,29}/ \IZ_3$} \label{sec:1.4}
This manifold can be obtained by contracting one of the last two $\IP^2$ spaces in the representation of $\mathscr{X}^{9,21}$  (see \sref{sec:1.2}). The configuration matrix is
\vskip 5pt
$$
\mathscr{X}^{8,29}~=~~
\cicy{\IP^1 \\ \IP^1\\  \vrule height10pt width0pt depth8pt  \IP^1\\ \IP^2\\ \IP^2\\ \vrule height10pt width0pt depth8pt  \IP^2\\ \IP^2}
{ \one& 0& 0&~ 0&0& 0&~ \one & 0\\
 0& \one& 0&~ 0&0& 0&~ \one & 0\\
\vrule height10pt width0pt depth8pt  0& 0& \one&~ 0&0& 0&~ \one & 0\\
\one& 0& 0&~ \one& 0& 0&~ 0 & \one\\
0& \one& 0&~ 0& \one& 0&~ 0 & \one\\
\vrule height10pt width0pt depth8pt  0& 0& \one&~0& 0& \one&~ 0 & \one\\
0& 0& 0&~\one&\one& \one&~ 0 & 0\\}_{-42}^{8,29}
$$
\vskip 5pt
with equivalent diagrams
\vspace{-21pt}
\begin{flushright}
$${\includegraphics[width=1.8in]{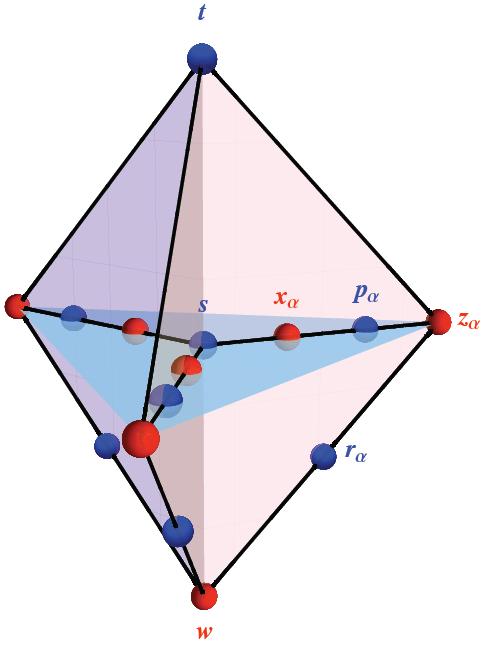}}
\hskip 52pt \lower -20pt\hbox{\includegraphics[width=2.4in]{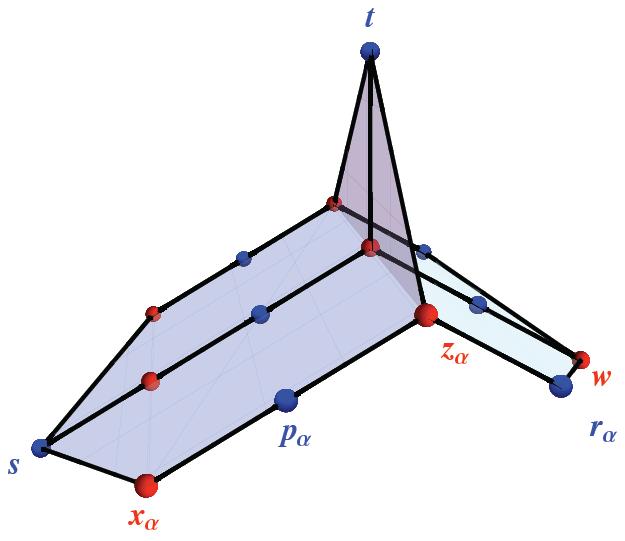}}
$$
\end{flushright}
\vspace{-8pt}
As before, take coordinates $x_{\a,j}$ for the three $\IP^1$ spaces, $z_{\a,a}$ for the first three $\IP^2$ spaces and $w_a$ for the last $\IP^2$ space. Denote by $p^\a$ the first three polynomials, by $r^\a$ the following three and by $s$ and $t$ the last two. The $\IZ_3$-action is generated by
\vspace{-20pt}
\begin{center}
\begin{minipage}{0.8\textwidth}
\begin{align*}
g:~x_{\alpha, j} & ~\to~  x_{\alpha+1, j}~,& z_{\alpha, a} &~\to~ z_{\alpha+1, a}~,& w_{a} &~\to~ w_{a+1}\\[5pt]
 p^{\alpha}&~\to~  p^{\alpha+1}~,& r &~\to~ r~, & s&~\to~ s~,~~~~~~~ t~\to~ t ~ . 
\end{align*}
\end{minipage}
\end{center}

The most general polynomials, covariant with the $g$ action 
\vspace{-10pt}
\begin{center}
\begin{minipage}{0.7\textwidth}
\begin{align*}
p^{\alpha} ~&=~ \sum_{j,a} P^\a_{ja}\, x_{\a, j}\, z_{\alpha, a}~, & r^{\a} ~&=~ \sum_{a,b} R^\a_{ab}\, z_{\a,a}w_{\a+b} \\[8pt]
s ~&=~ \sum_{i,j,k} S_{ijk}\, x_{0,i}\, x_{1,j}\, x_{2,k}~, & t ~&=~ \sum_{a,b,c} T_{abc}\, z_{0,a}\, z_{1,b}\, z_{2,c}
\end{align*}
\end{minipage}
\end{center}

where $S_{ijk}$ and $T_{abc}$ are cyclic in their indices and, for given $a,b,j$, the coefficients 
$P^\a_{ja}, R^\a_{ab}$ take the same values for all $\a$'s. It is easy to show that the action is fixed point free. 

A parameter count shows that there are $80$ terms in the most general form of the polynomials, before imposing covariance under the $\IZ_3$ action. Of these, $41$ parameters can be eliminated by coordinate redefinitions, $8$ by rescalings of the polynomials and $2$ more by redefinitions of $t$ in terms of $D_1 = \det \left.\partial\left(p; x\right) \right|_z$ and $D_2 = \det \left.\partial\left(r; w\right) \right|_z$. The parameter count confirms the value of $h^{2,1} = 29 = 80 - 41 - 10$. 

After the $\IZ_3$ identification, there are $30$ possible terms in the polynomials, $13$ coordinate redefinitions and $6$ degrees of freedom associated with redefinitions of polynomials. These give a total of $11$ free parameters, thus $h^{2,1} = 11$ for the quotient manifold. 

In order to find $h^{1,1}$, let us consider an extended representation of the manifold, obtained by splitting $s$ by means of a $\IP^2$, as shown in Table \ref{tab(8,29)Extended}. We see that the quotient 
manifold~has~\hbox{$h^{1,1} = 4$}. 
\vskip 20pt

\begin{table}[ht]
\begin{center}
\boxed{\begin{tabular}{ccc}
& & $\ $ \\
\hskip .3in $\cicy{\IP^1 \\ \IP^1\\ \vrule height10pt width0pt depth8pt  \IP^1\\ \IP^2\\ \IP^2\\ \vrule height10pt width0pt depth8pt  \IP^2\\ \IP^2\\ \IP^2}
{ \one& 0& 0& ~0& 0& 0&~\one& 0& 0&~ 0 \\
 0& \one& 0& ~0& 0&0&~0& \one& 0&~ 0 \\
\vrule height10pt width0pt depth8pt  0& 0& \one&~ 0& 0&0&~0&0& \one&~ 0 \\
\one & 0& 0&~ \one& 0&0&~0& 0& 0&~ \one \\
 0& \one & 0&~ 0& \one& 0&~ 0& 0& 0&~ \one \\
 \vrule height10pt width0pt depth8pt 0& 0& \one&~ 0& 0& \one &~0&0& 0&~ \one \\
0& 0& 0 &~ \one& \one &\one &~ 0& 0& 0&~0 \\
0& 0& 0&~ 0& 0&0&~\one& \one& \one&~ 0 \\}_{-42}^{8,29}
$ &  \hskip .3in \lower 70pt \hbox{\includegraphics[width=2.4in]{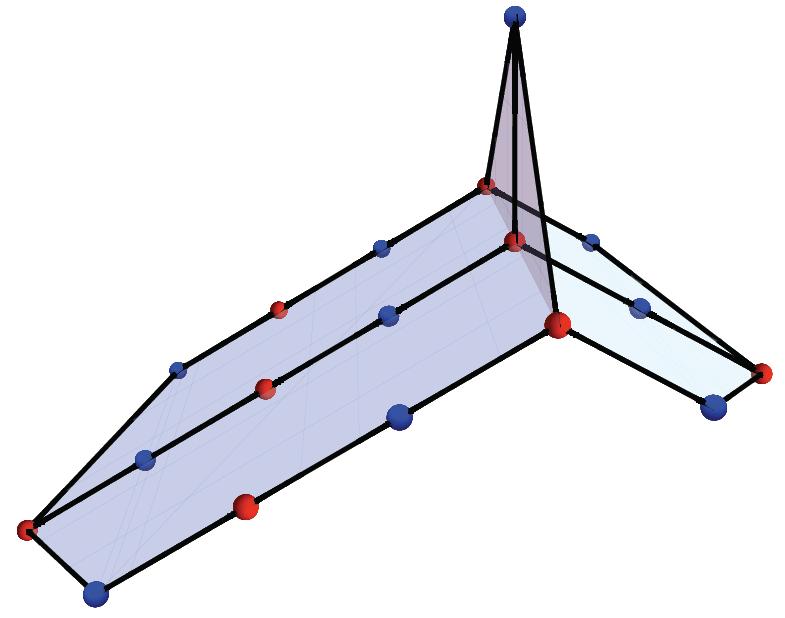}}& $\ $\\
 & & $\ $ 
\end{tabular}}
\capt{5.5in}{tab(8,29)Extended}{The matrix and diagram for the extended representation of 
$\mathscr{X}^{8,29}$ for which all 8 K\"ahler forms are represented by ambient spaces. After the $\IZ_3$ identification, only $4$ of the $(1,1)$--forms remain independent.} 
\end{center}
\end{table}

We have checked that the polynomials are transverse. Thus we obtain a smooth quotient of a CICY manifold  $\mathscr{X}^{4,11} = \mathscr{X}^{8,29}/ \IZ_3$  with fundamental group $\IZ_3$ and $\hodgenos = (4,11)$.


\subsection{The manifold $\mathscr{X}^{7, 37}$ with quotient
 $\mathscr{X}^{3, 13} =  \mathscr{X}^{7,37}/ \IZ_3$} \label{sec:1.5}
\vspace{0pt}
This manifold can be obtained by contracting the last $\IP^2$ space of the previous 
manifold, $\mathscr{X}^{8,29}$. The configuration matrix is
$$
\mathscr{X}^{7,37}~=~~
\cicy{\IP^1 \\ \IP^1\\  \vrule height10pt width0pt depth8pt  \IP^1\\ \IP^2\\ \IP^2\\ \vrule height10pt width0pt depth8pt  \IP^2}
{ \one& 0& 0&~ \one & 0 & 0\\
 0& \one& 0&~  \one & 0 & 0\\
\vrule height10pt width0pt depth8pt  0& 0& \one&~ \one & 0 & 0\\
\one& 0& 0&~  0 & \one & \one\\
0& \one& 0&~  0 & \one & \one\\
\vrule height10pt width0pt depth8pt  0& 0& \one&~ 0 & \one & \one}_{-60}^{7,37}
$$
\vskip 5pt
with equivalent diagrams
\vspace{-15pt}
\begin{flushright}
$${\includegraphics[width=2.1in]{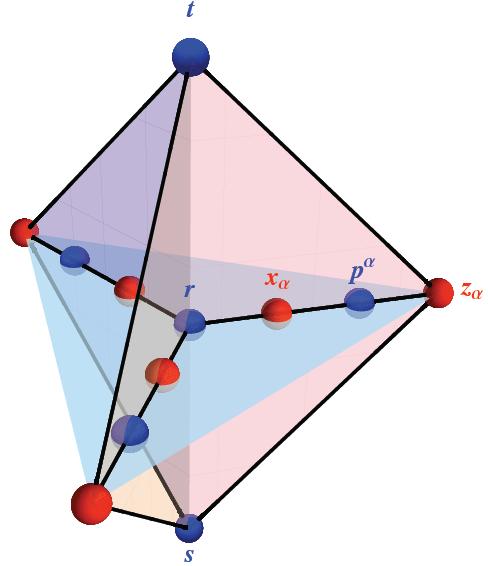}}
\hskip 52pt \lower 0pt\hbox{\includegraphics[width=2.4in]{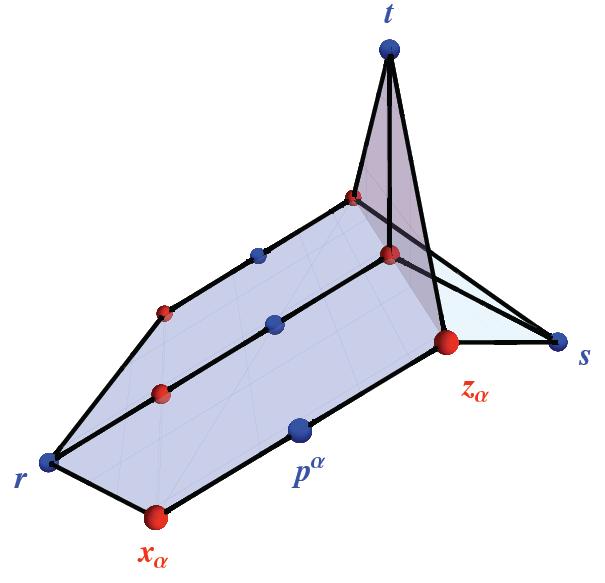}}
$$
\end{flushright}

As before, take coordinates $x_{\a,j}$ for the three $\IP^1$ spaces and $z_{\a,a}$ for the three $\IP^2$'s. Denote by $p^\a$ the first three polynomials, and by $r,s,t$  the last three. The $\IZ_3$ action is generated by 
\vspace{-30pt}
\begin{center}
\begin{minipage}{0.8\textwidth}
\begin{align*}
g:~x_{\alpha, j} & ~\to~  x_{\alpha+1, j}~,& z_{\alpha, a} &~\to~ z_{\alpha+1, a}\\[5pt]
 p^{\alpha}&~\to~  p^{\alpha+1}~,& r &~\to~ r~, & s~\to~ s~,~~~~~~~~t~\to~ t ~ . 
\end{align*}
\end{minipage}
\end{center}

The most general polynomials, covariant with the $g$ action are
\vspace{-10pt}
\begin{center}
\begin{minipage}{0.7\textwidth}
\begin{align*}
p^{\alpha} ~&=~ \sum_{j,a} P^\a_{ja}\, x_{\a, j}\, z_{\alpha, a}~,& r~&=~ \sum_{i,j,k} R_{ijk}\, x_{0,i}\, x_{1,j}\, x_{2,k}\\[8pt]
s ~&=~ \sum_{a,b,c} S_{abc}\, z_{0,a}\, z_{1,b}\, z_{2,c}~, & t ~&=~ \sum_{a,b,c} T_{abc}\, z_{0,a}\, z_{1,b}\, z_{2,c}
\end{align*}
\end{minipage}
\end{center}
where $R_{ijk}$, $S_{abc}$ and $T_{abc}$ are cyclic in their indices and the coefficients $P^\a_{ja}$ do not depend on $\a$'s. The search for fixed points is very similar to the previous cases and gives an empty result.

The parameter count shows that there are again $80$ terms in the most general form of the polynomials, before imposing covariance under the $\IZ_3$ action. Of these, $33$ parameters can be eliminated by coordinate redefinitions, $4$ by rescaling the polynomials $r$ and $p^\a$. The most general redefinitions for $t$ and $s$ involve $6$ parameters: 
$$ t~ \to~ c_{tt} t + c_{ts} s + c_{1} D~, ~~~~s ~\to~ c_{st} t + c_{ss} s + c_{2} D$$
where $D = \det \left.\partial\left(p; x\right) \right|_z$. Thus the number of free parameters is $80-33-10 = 37$, confirming the value of $h^{2,1}$.

In order to find $h^{1,1}$, let us consider an extended representation of the manifold, obtained by splitting $s$ with a $\IP^2$, as shown in Table \ref{tab(7,37)Extended}. It follows that the quotient manifold has 
$h^{1,1} = 3$. Thus, we obtain a quotient $\mathscr{X}^{3,13} = \mathscr{X}^{7,37}/ \IZ_3$.

\begin{table}
\begin{center}
\boxed{\begin{tabular}{ccc}
& & $\ \ $\\
\hskip .3in $\cicy{\IP^1 \\ \IP^1\\ \vrule height10pt width0pt depth8pt  \IP^1\\ \IP^2\\ \IP^2\\ \vrule height10pt width0pt depth8pt  \IP^2\\ \IP^2}
{ \one& 0& 0& ~\one& 0& 0&~ 0 & 0\\
 0& \one& 0&~0& \one& 0&~ 0 & 0\\
\vrule height10pt width0pt depth8pt  0& 0& \one&~0&0& \one&~ 0 & 0\\
\one & 0& 0&~0& 0& 0&~ \one & \one\\
 0& \one & 0&~ 0& 0& 0&~ \one & \one \\
 \vrule height10pt width0pt depth8pt 0& 0& \one&~0&0& 0&~ \one & \one \\
0& 0& 0&~\one& \one& \one&~ 0 & 0\\}_{-60}^{7,37}
$ &  \hskip .5in \lower 75pt \hbox{\includegraphics[width=2.4in]{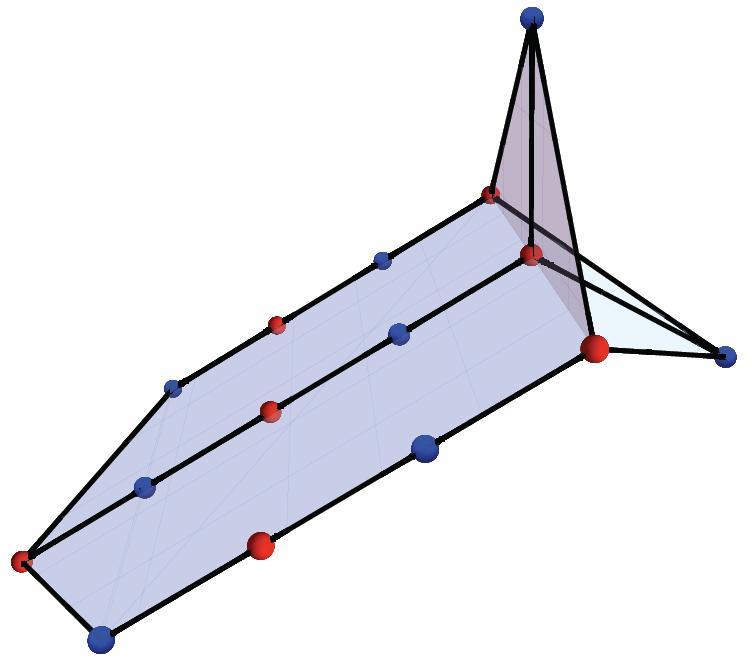}} & $\ $\\
& & $\ $
\end{tabular}}
\capt{5.5in}{tab(7,37)Extended}{The matrix and diagram for the extended representation of 
$\mathscr{X}^{7,37}$ for which all 7 K\"ahler forms are represented by ambient spaces. After the $\IZ_3$ identification, only $3$ of the $(1,1)$--forms remain independent.} 
\end{center}
\end{table}

We have checked that the polynomials are transverse. Thus we obtain a smooth Calabi-Yau manifold with fundamental group $\IZ_3$ and $\hodgenos = (3,13)$.


\subsection{The manifold $\mathscr{X}^{11,29}$ with quotient
$\mathscr{X}^{5, 11} = \mathscr{X}^{11,29}/ \IZ_3$} \label{sec:1.7}
\vskip -8pt
The configuration matrix for this manifold is 
$$
\mathscr{X}^{11,29}~=~~
\cicy{\IP^1 \\ \IP^1\\  \vrule height10pt width0pt depth8pt  \IP^1\\ \IP^3\\ \IP^2}
{ \one& \one&~ 0 & 0 & 0\\
 \one& \one& ~ 0 & 0 & 0\\
\vrule height10pt width0pt depth8pt  \one& \one&~ 0 & 0 & 0\\
0 &\one& ~ \one & \one & \one\\
0& 0&~  \one & \one & \one}_{-36}^{11,29}
\hskip0.35in
\lower0.34in\hbox{\includegraphics[width=2.5in]{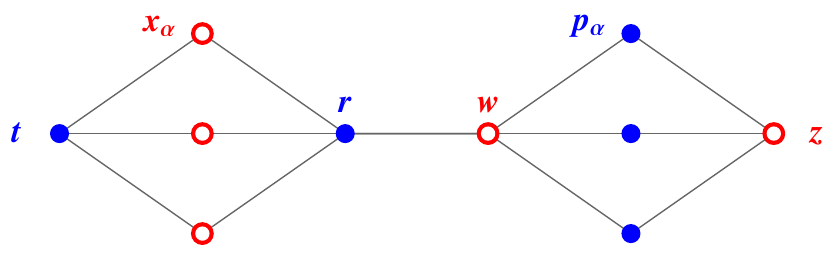}}
$$

Let $x_{\a,j}$ denote the homogeneous coordinates of the three $\IP^1$ spaces, $\left( w_0, w_a\right)$ those of the $\IP^3$ and $z_a$ the coordinates of  the $\IP^2$. As before, $\a$ and $a$ take values in $\IZ_3$ and $j$ in $\IZ_2$.  Also, denote by $t$ and $r$ the first two polynomials and by $p^\a$ the last three. 

We can write the most general form of the defining polynomials as
\begin{equation*}
\begin{split}
&t = \sum_{i,j,k} T_{ijk}\, x_{0,i}\, x_{1,j}\, x_{2,k}\\[5pt]
& r = \sum_{a, i, j, k} R_{a i j k} \, x_{0,i}\, x_{1,j}\, x_{2,k}\, w_a + w_0\sum_{ i, j, k} \widetilde{R}_{i j k} \, x_{0,i}\, x_{1,j}\, x_{2,k}\\[5pt]
& p^\a = \sum_{\a, a,b} P^\a_{ab} z_{\a + a} w_{\a + b} + w_0 \sum_{\a, a} \widetilde{P}^\a_a z_{\a+a}\\
\end{split}
\end{equation*}
There are 76 terms in the polynomials, out of which 32 can be eliminated by redefinitions of coordinates, 1 by rescaling $t$, 5 by rescaling $r$ and adding to it polynomials of the form $C_q\, w_q\, t$ ($q\in \IZ_4$) and 9 by redefining the polynomials $p^{\a}$. This leaves 29 free parameters, which agree with the number of complex structure parameters. Note, however, that in this case we cannot rely on our previous method for computing $h^{2,1}$ for the quotient manifold, since the diagram associated with ${\mathscr X}^{11,29}$ does not fulfil the sufficient condition discussed in the introduction. Assuming, however, that the method is still valid, we can compute a value for $h^{2,1}\left( {\mathscr X}^{11,29} /\IZ_3\right)$ and compare it against the value obtained by expressing $h^{2,1}$ in terms of the Euler characteristic and $h^{1,1}$. 

The $\IZ_3$ action is generated by 
\vspace{-20pt}
\begin{center}
\begin{minipage}{0.8\textwidth}
\begin{align*}
g:~x_{\alpha, j} & ~\to~  x_{\alpha+1, j}~,& z_{\alpha} &~\to~ z_{\alpha+1} ~, &\left( w_0, w_{a}\right) & ~\to~ \left( w_0, w_{a+1}\right)\\[5pt]
 p^{\alpha}&~\to~  p^{\alpha+1}~,& r &~\to~ r~, & t & ~\to~ t ~ . 
\end{align*}
\end{minipage}
\end{center}
\vspace{-5pt}
The polynomials are covariant under this action if $T_{ijk}$ and $\widetilde{R}_{ijk}$ are cyclic in their indices, $R_{aijk} = R_{a+1, kji} = R_{a+2, jik}$ and $P^\a_{ab}, \widetilde{P}^\a_a$ do not depend on $\a$. This leaves a total of 28 terms in the polynomials, out of which 10 can be eliminated by coordinate redefinitions and 7 by redefinitions of the $\IZ_3$-covariant polynomials. The parameter count indicates that there are $28-17 =11$ complex structure parameters, a result which is consistent with the value $h^{1,1}\left( {\mathscr X}^{11,29} /\IZ_3 \right) = 5$, obtained from the extended representation in Table \ref{tab(11,29)Extended}. 

Fixed points of $g$, in the embedding space, are of the form $x_{\a,j} = x_j$, and $z_\a = \xi^\a$, $w_a = \zeta^\a$, where $\xi^\a$ and $\zeta^\a$ are cube roots of unity. The last three polynomials also determine the value of $w_0$, leaving two equations $t\left(x_j\right) = r\left(x_j\right) = 0$ which have no solutions in $\IP^1$, in general. Therefore, the action is fixed point free, on the manifold.

\begin{table}[H]
\begin{center}
\boxed{\begin{tabular}{cccc}
& & & \\
$\ $ & \hskip .0in \footnotesize $\cicy{\IP^1 \\ \IP^1\\ \vrule height10pt width0pt depth8pt   \IP^1\\ \IP^1\\ \IP^1\\ \vrule height10pt width0pt depth8pt  \IP^1\\ \IP^1\\ \IP^1\\ \vrule height10pt width0pt depth8pt  \IP^1\\  \IP^2\\ \IP^3}
{ \one& 0& 0&~ \one&0& 0&~ 0 & 0 & 0&~0& 0  \\
 0& \one& 0&~ 0& \one&  0&~ 0 & 0 & 0&~0 & 0\\
\vrule height10pt width0pt depth8pt  0& 0& \one&~ 0 & 0& \one&~ 0 & 0 & 0 &~0 & 0\\
  \one& 0&0& ~ 0& 0& 0&~ \one & 0 & 0&~0 & 0\\
  0& \one& 0&~ 0& 0& 0&~ 0 & \one & 0&~0 & 0\\
\vrule height10pt width0pt depth8pt    0& 0& \one &~0& 0& 0&~ 0 & 0 & \one &~0 & 0\\
0& 0& 0&~0& 0& 0&~0& 0& 0&~ \one & \one\\
0& 0& 0&~0& 0& 0&~0& 0& 0&~\one & \one\\
\vrule height10pt width0pt depth8pt  0& 0& 0&~0& 0& 0&~0& 0& 0&~ \one & \one\\
0& 0& 0&~\one&\one& \one&~0&0& 0 &~0 & 0\\
 0& 0& 0&~ 0& 0 & 0&~\one&\one& \one& ~\one& 0}_{-36}^{11,29}
$ &  \hskip-0.3in \lower21pt \hbox{\includegraphics[width=3.3in]{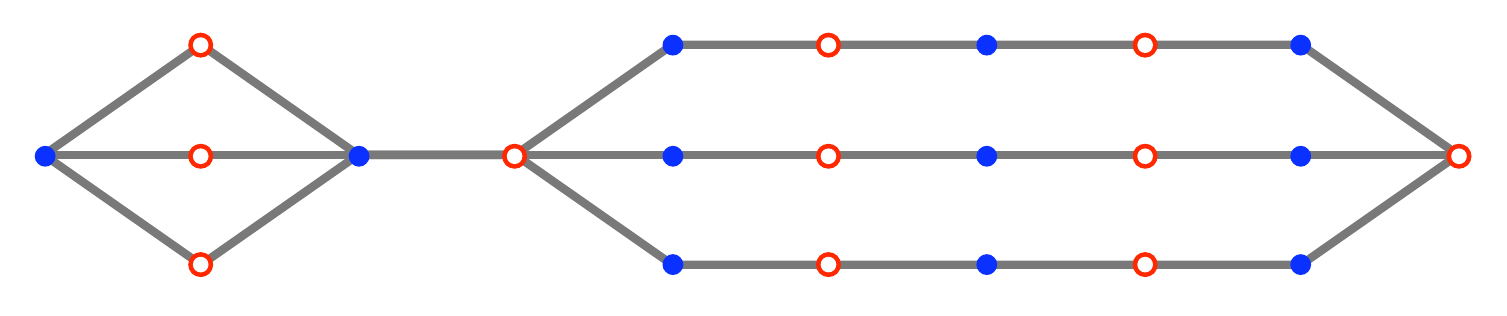}}\hspace{-0.2in} &\\
& & &  
\end{tabular}}
\capt{5.5in}{tab(11,29)Extended}{The matrix and diagram for the extended representation of $\mathscr{X}^{11,29}$ for which all 11 K\"ahler forms are represented by ambient spaces. After the $\IZ_3$ identification, only $5$ of the $(1,1)$--forms remain independent.} 
\end{center}
\end{table}

There is another manifold for which the list \cite{Volker} indicates a $\IZ_3$-free symmetry, namely the manifold, or rather the deformation class, given by the following configuration matrix and diagram: 
$$
\mathscr{X}^{11,29}~=~~
\cicy{\IP^2 \\   \IP^2\\ \IP^3\\ \IP^2}
{ \one& \one & \one&~ 0 & 0 & 0\\
 \one& \one& \one &  ~ 0 & 0 & 0\\
0& 0 &\one& ~ \one & \one & \one\\
0& 0& 0&~  \one & \one & \one}_{-36}^{11,29}
\hskip0.25in
\lower0.4in\hbox{\includegraphics[width=2.5in]{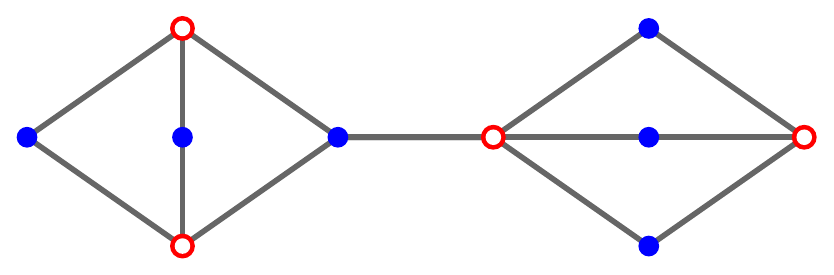}}
$$

Note, however, that this is only an alternative representation of the previous manifold, since
$$
\cicy{\IP^2 \\   \IP^2}
{ \one& \one\\
 \one& \one} \,\cong~
\cicy{\IP^1 \\   \IP^1\\ \IP^1}
{ \one\\ \one \\ \one} 
$$
are equivalent alternative representations of the del Pezzo surface $\mathrm{dP}_6$.


\newpage
\section{The web of $\IZ_3$ quotients}\label{secWeb}
\vskip -8pt
CICY threefolds form a web, connected by conifold transitions. In \cite{CD08}, it was suggested that a similar statement might hold for the webs of discrete quotients of CICY threefolds. The purpose of this section is to clarify this statement. 
Since conifold transitions do not affect the fundamental group, there will be several webs corresponding to different fundamental groups. Here, we are mainly concerned with the web of smooth $\IZ_3$ quotients of CICY threefolds, having fundamental group $\IZ_3$. 

\subsection{Conifold transitions between smooth quotients of Calabi-Yau threefolds}
Let $\Xcheck \to \Xsharp \to \mathscr{X}$ be a conifold transition between the 
\cy threefolds $\Xcheck$ and $\mathscr{X}$, where $\chi\left( {\Xcheck} \right) > \chi ( \mathscr{X} )$ and 
$\Xsharp$ is a conifold. Assume also that $\Xcheck$ and $\mathscr{X}$ both admit a free action of $G$. 

The argument\footnote{This argument is partly due to Rhys Davies.} presented below shows that, if the two actions of $G$ agree on $\Xsharp$, then the conifold transition commutes with taking quotients:
\beqnn
\begin{CD}
 \Xcheck   @>\wp >> \Xsharp @>\mathrm{deformation}>> \mathscr{X}\\
@V \mathrm{~} VV @VV \mathrm{~}V@VV \mathrm{~}V\\
\Xcheck / G @> \bar \wp  >> \Xsharp / G @>\mathrm{deformation}>> \mathscr{X} / G
\end{CD}
\eeqnn
where the map $\wp$ projects an entire $\IP^1\subset {\Xcheck}$ to each node of $\Xsharp$ and the deformation replaces each node of the conifold with a copy of $S^3$.  

The free $G$ action on $\Xcheck$ descends to a free action of $G$ on $\Xsharp$. Indeed, the conifold inherits a $G$-action from ${\Xcheck}$ via the projection $\wp: {\Xcheck} \to \Xsharp$. Moreover, the action of $G$ on the conifold is free, as we can see by the following argument. The only points which could be fixed by the group action are the conifold singularities. Assume this was the case for a node $p \in \Xsharp$. Then there exists an element 
$g\in G$ such that $g(p) = p$. But the $G$-action on $\Xsharp$ descends from the free $G$-action on $\Xcheck$, where $p$ is replaced by an $S^2$, and hence this implies the existence of a holomorphic map $g:S^2 \to S^2$, which being free, can have no fixed points, but there is no such holomorphic map. 
Thus, the action permutes the nodes of the conifold. In particular, the number of nodes of the conifold quotient $\Xsharp/ G$ reduces by a factor equal to the order of $G$. 

For all the quotients of CICY threefolds discussed above, in \cite{CD08} and \cite{Triadophilia}, the group actions are given explicitly in terms of homogeneous coordinates of the embedding spaces. This makes it very easy to decide whether, in a conifold transition, the actions on the resolved spaces descend to the same action on the conifold. 
\subsection{The web of $\IZ_3$ quotients	}
\vskip -8pt
The manifolds discussed in the previous section complete the web of smooth $\IZ_3$-free quotients of CICY threefolds in a nice way. Five of the new quotients can be organized in the double sequence that we have already seen in the introduction, but which we repeat here for completeness. The arrows of the table stand for conifold transitions, and the appended numbers are the respective Hodge numbers: 
\begin{table}[ht]
\centering
\capt{6.0in}{doublesequenceBody}{The first double sequence of $\IZ_3$-free CICY quotients. In red, the new quotients.}
\vskip8pt
\framebox[6.0in]{\centering
\parbox{5.2in}{\vspace{4pt}
For the covering spaces:   
$\Delta_{\raisebox{-3pt}{$\scriptstyle{}\hskip-2pt\color{blue}\rightarrow $}} \hodgenos =\left(3, -3\right)
\,;~\Delta_{\color{blue}\downarrow}  \hodgenos = \left( 1, -8 \right) $ \\[4pt]
For the quotient spaces: 
$\Delta_{\raisebox{-3pt}{$\scriptstyle{}\hskip-2pt\color{blue}\rightarrow $}} \hodgenos =\left(1, -1\right)
\,;~\Delta_{\color{blue}\downarrow}  \hodgenos = \left( 1, -2 \right) $
\vspace{-8pt}} }
\vskip8pt
\framebox[6.0in]{
\begin{tabular}{c}
\begin{tikzpicture}[scale=1.2]
\clip (-2.5, -.4) rectangle (9.55,5.7);
\def\nodeshadowed[#1]#2;{\node[scale=1.1,above,#1]{#2};}

\nodeshadowed [at={(-1,0 )},yslant=0.0]
{ {\small \textcolor{black} {$\mathbf{\left( \mathscr{X}^{6,24}/ \IZ_3 \right) ^{4,10}}$}} };
\nodeshadowed [at={(2,0 )},yslant=0.0]
{ {\small \textcolor{black} {$\mathbf{\color{red} \left( \mathscr{X}^{9,21}/ \IZ_3 \right) ^{5,9}}$ }} };
\nodeshadowed [at={(5,0 )},yslant=0.0]
{ {\small \textcolor{black} {$\mathbf{\color{red} \left( \mathscr{X}^{12,18}/ \IZ_3 \right) ^{6,8}}$ }} };
\nodeshadowed [at={(8,0 )},yslant=0.0]
{ {\small \textcolor{black} {$\mathbf{\left( \mathscr{X}^{15,15}/ \IZ_3 \right) ^{7,7}}$}} };
\nodeshadowed [at={(-1,1.5 )},yslant=0.0]
{ {\small \textcolor{black} {$\mathbf{\left( \mathscr{X}^{5,32}/ \IZ_3 \right) ^{3,12}}$ }} };
\nodeshadowed [at={(2,1.5 )},yslant=0.0]
{ {\small \textcolor{black} {$\mathbf{\color{red}  \left( \mathscr{X}^{8,29}/ \IZ_3 \right) ^{4,11}}$}} };
\nodeshadowed [at={(5,1.5 )},yslant=0.0]
{ {\small \textcolor{black} {$\mathbf{\color{red} \left( \mathscr{X}^{11,26}/ \IZ_3 \right) ^{5,10}}$}} };
\nodeshadowed [at={(-1,3 )},yslant=0.0]
{ {\small \textcolor{black} {$\mathbf{\left( \mathscr{X}^{4,40}/ \IZ_3 \right) ^{2,14}}$}} };
\nodeshadowed [at={(2,3 )},yslant=0.0]
{ {\small \textcolor{black} {$\mathbf{\color{red} \left(  \mathscr{X}^{7,37} / \IZ_3 \right) ^{3,13}}$}} };
\nodeshadowed [at={(-1,4.5 )},yslant=0.0]
{ {\small \textcolor{black} {$\mathbf{\left( \mathscr{X}^{3,48}/ \IZ_3 \right) ^{1,16}}$}} };

\draw[very thick,blue,->] (-1, 4.5) -- (-1,3.7);
\draw[very thick,blue,->] (-1, 3.) -- (-1,2.2);
\draw[very thick,blue,->] (-1, 1.5) -- (-1,.7);
\draw[very thick,blue,->] (2, 3.) -- (2,2.2);
\draw[very thick,blue,->] (2, 1.5) -- (2,.7);
\draw[very thick,blue,->] (5, 1.5) -- (5,.7);

\draw[very thick,blue,->] (-.1, 3.3) -- (.8, 3.3);
\draw[very thick,blue,->] (-.1, 1.8) -- (.8, 1.8);
\draw[very thick,blue,->] (-.1, .3) -- (.8, .3);
\draw[very thick,blue,->] (2.9, 1.8) -- (3.74, 1.8);
\draw[very thick,blue,->] (2.9, .3) -- (3.74, .3);
\draw[very thick,blue,->] (5.9, .3) -- (6.8, .3);

\end{tikzpicture}
\end{tabular}}
\end{table}

Note that each horizontal arrow corresponds to the same change in the Hodge numbers, as does each vertical arrow. This is also the case for the Hodge numbers of the quotient manifolds in the following sequence, to which the sixth new manifold belongs: 

\begin{center}
\begin{table}[H]
\centering
\capt{6.0in}{OneLineSequence}{Sequence of $\IZ_3$-free CICY quotients. In red, one of the new quotients.}
\vskip8pt
\framebox[6.0in]{\centering
\parbox{5.2in}{\centering\vspace{2pt}
For the quotient spaces: 
$\Delta_{\raisebox{-3pt}{$\scriptstyle{}\hskip-2pt\color{blue}\rightarrow $}} \hodgenos =\left(1, -2\right)$
\vspace{0pt}} }
\vskip8pt
\framebox[6.0in]{
\begin{tabular}{c}
\begin{tikzpicture}[scale=1.2]
\clip (-2.5, -.2) rectangle (9.55,1.1);
\def\nodeshadowed[#1]#2;{\node[scale=1.1,above,#1]{#2};}

\nodeshadowed [at={(-1,0 )},yslant=0.0]
{ {\small \textcolor{black} {$\mathbf{\left( \mathscr{X}^{8,35}/\IZ_3 \right)^{4,13} }$}} };
\nodeshadowed [at={(2,0 )},yslant=0.0]
{ {\small \textcolor{black} {$\mathbf{\color{red} \left( \mathscr{X}^{11,29}/\IZ_3 \right)^{ 5,11} }$}} };
\nodeshadowed [at={(5,0 )},yslant=0.0]
{ {\small \textcolor{black} {$\mathbf{\left( \mathscr{X}^{14,23}/\IZ_3 \right)^{ 6,9}}$ }} };
\nodeshadowed [at={(8,0 )},yslant=0.0]
{ {\small \textcolor{black} {$\mathbf{\left( \mathscr{X}^{19,19} /\IZ_3 \right)^{7,7} }$}} };

\draw[very thick,blue,->] (-.1, .3) -- (.74, .3);
\draw[very thick,blue,->] (2.9, .3) -- (3.74, .3);
\draw[very thick,blue,->] (5.9, .3) -- (6.77, .3);

\end{tikzpicture}
\end{tabular}}
\end{table}
\end{center}

There is yet a third sub-web of smooth $\IZ_3$-quotients which can be organized into a double sequence: 

\begin{table}[ht]
\centering
\capt{6.0in}{doublesequenceAppendix}{The second double sequence of $\IZ_3$-free CICY quotients.}
\vskip8pt
\framebox[6.0in]{\centering
\parbox{5.2in}{\vspace{4pt}
For the covering spaces:   
$\Delta_{\raisebox{-3pt}{$\scriptstyle{}\hskip-2pt\color{blue}\rightarrow $}} \hodgenos =\left(3, -6\right)
\,;~\Delta_{\color{blue}\downarrow}  \hodgenos = \left( 1, -7 \right) $ \\[4pt]
For the quotient spaces: 
$\Delta_{\raisebox{-3pt}{$\scriptstyle{}\hskip-2pt\color{blue}\rightarrow $}} \hodgenos =\left(1, -2\right)
\,;~\Delta_{\color{blue}\downarrow}  \hodgenos = \left( 1, -5 \right) $
\vspace{-8pt}} }
\vskip8pt
\framebox[6.0in]{
\begin{tabular}{c}
\begin{tikzpicture}[scale=1.2]
\clip (-1.3, 1.5) rectangle (8.35,5.4);
\def\nodeshadowed[#1]#2;{\node[scale=1.1,above,#1]{#2};}

\nodeshadowed [at={(.5,1.5 )},yslant=0.0]
{ {\small \textcolor{black} {$\mathbf{\left( \mathscr{X}^{3,39} / \IZ_3 \right) ^{ 3,15}} $}} };
\nodeshadowed [at={(3.5,1.5 )},yslant=0.0]
{ {\small \textcolor{black} {$\mathbf{\left( \mathscr{X}^{6,33} / \IZ_3 \right) ^{4,13}} $}} };
\nodeshadowed [at={(6.5,1.5 )},yslant=0.0]
{ {\small \textcolor{black} {$\mathbf{\left( \mathscr{X}^{9,27}/ \IZ_3 \right) ^{5,11}} $}} };
\nodeshadowed [at={(.5,3 )},yslant=0.0]
{ {\small \textcolor{black} {$\mathbf{\left( \mathscr{X}^{2,56} / \IZ_3 \right) ^{ 2,20}}$}} };
\nodeshadowed [at={(3.5,3 )},yslant=0.0]
{ {\small \textcolor{black} {$\mathbf{\left( \mathscr{X}^{5,50} / \IZ_3 \right) ^{ 3,18}} $}} };
\nodeshadowed [at={(.5,4.5 )},yslant=0.0]
{ {\small \textcolor{black} {$\mathbf{\left( \mathscr{X}^{1,73}/ \IZ_3 \right) ^{ 1,25}}$}} };

\draw[very thick,blue,->] (.5, 4.5) -- (.5,3.7);
\draw[very thick,blue,->] (.5, 3.) -- (.5,2.2);
\draw[very thick,blue,->] (3.5, 3.) -- (3.5,2.2);

\draw[very thick,blue,->] (1.4, 3.3) -- (2.3, 3.3);
\draw[very thick,blue,->] (1.4, 1.8) -- (2.3, 1.8);
\draw[very thick,blue,->] (4.4, 1.8) -- (5.24, 1.8);

\end{tikzpicture}
\end{tabular}}
\end{table}
\vspace{10pt}
The last two sequences above partially overlay each other in \fref{Z3Web}. However, they do not form a single structure as is seen by examining the Hodge numbers of the covering manifolds.

The sequences above are part of the connected web of conifold transitions between all the $\IZ_3$-free quotients of CICY threefolds, shown in Table \ref{TheWeb}. In order to save space, the chart contains only the Hodge numbers. For example, $\mathbf{\frac{3,48}{1,16}}~$ should be read as 
$\mathbf{\left( \mathscr{X}^{3,48}/ \IZ_3 \right) ^{1,16}}$.

\begin{center}
\begin{table}[H]
\begin{center}
\capt{6.0in}{TheWeb}{The web of $\IZ_3$-free CICY quotients.}
\vskip8pt
\framebox[6.0in]{
\begin{tabular}{c}
\begin{tikzpicture}[scale=.71]
\clip (0,-.3) rectangle (19.4,8.4);
\def\nodeshadowed[#1]#2;{\node[scale=.94,above,#1]{#2};}

\nodeshadowed [at={(.5,6 )},yslant=0.0]
{ {\small \textcolor{black} {$\mathbf{\dfrac{3,48}{1,16}}$}} };
\nodeshadowed [at={(2.5,6 )},yslant=0.0]
{ {\small \textcolor{black} {$\mathbf{\dfrac{4,40}{2,14}}$}} };
\nodeshadowed [at={(4.5,6 )},yslant=0.0]
{ {\small \textcolor{black} {$\mathbf{\dfrac{5,32}{3,12}}$}} };
\nodeshadowed [at={(6.5,6 )},yslant=0.0]
{ {\small \textcolor{black} {$\mathbf{\dfrac{6,24}{4,10}}$}} };
\nodeshadowed [at={(2.5,4 )},yslant=0.0]
{ {\small \textcolor{red} {$\mathbf{\dfrac{7,37}{3,13}}$}} };
\nodeshadowed [at={(4.5,4 )},yslant=0.0]
{ {\small \textcolor{red} {$\mathbf{\dfrac{8,29}{4,11}}$}} };
\nodeshadowed [at={(6.5,4 )},yslant=0.0]
{ {\small \textcolor{red} {$\mathbf{\dfrac{9,21}{5,9}}$}} };
\nodeshadowed [at={(4.5,2 )},yslant=0.0]
{ {\small \textcolor{red} {$\mathbf{\dfrac{11,26}{5,10}}$}} };
\nodeshadowed [at={(6.5,2 )},yslant=0.0]
{ {\small \textcolor{red} {$\mathbf{\dfrac{12,18}{6,8}}$}} };
\nodeshadowed [at={(6.5,0 )},yslant=0.0]
{ {\small \textcolor{black} {$\mathbf{\dfrac{15,15}{7,7}}$}} };

\nodeshadowed [at={(8.5,6 )},yslant=0.0]
{ {\small \textcolor{black} {$\mathbf{\dfrac{3,48}{3,18}}$}} };
\nodeshadowed [at={(8.5,4 )},yslant=0.0]
{ {\small \textcolor{black} {$\mathbf{\dfrac{6,33}{4,13}}$}} };
\nodeshadowed [at={(8.5,2 )},yslant=0.0]
{ {\small \textcolor{black} {$\mathbf{\dfrac{9,21}{5,9}}$}} };

\nodeshadowed [at={(10.5,6 )},yslant=0.0]
{ {\small \textcolor{black} {$\mathbf{\dfrac{8,35}{4,13}}$}} };
\nodeshadowed [at={(10.5,4 )},yslant=0.0]
{ {\small \textcolor{red} {$\mathbf{\dfrac{11,29}{5,11}}$}} };
\nodeshadowed [at={(10.5,2 )},yslant=0.0]
{ {\small \textcolor{black} {$\mathbf{\dfrac{14,23}{6,9}}$}} };
\nodeshadowed [at={(10.5,0 )},yslant=0.0]
{ {\small \textcolor{black} {$\mathbf{\dfrac{19,19}{7,7}}$}} };

\nodeshadowed [at={(12.5,6 )},yslant=0.0]
{ {\small \textcolor{black} {$\mathbf{\dfrac{2,83}{2,29}}$}} };
\nodeshadowed [at={(12.5,4 )},yslant=0.0]
{ {\small \textcolor{black} {$\mathbf{\dfrac{5,59}{3,21}}$}} };
\nodeshadowed [at={(12.5,2 )},yslant=0.0]
{ {\small \textcolor{black} {$\mathbf{\dfrac{8,44}{4,16}}$}} };

\nodeshadowed [at={(14.5,6 )},yslant=0.0]
{ {\small \textcolor{black} {$\mathbf{\dfrac{3,39}{3,15}}$}} };
\nodeshadowed [at={(16.5,6 )},yslant=0.0]
{ {\small \textcolor{black} {$\mathbf{\dfrac{2,56}{2,20}}$}} };
\nodeshadowed [at={(18.5,6 )},yslant=0.0]
{ {\small \textcolor{black} {$\mathbf{\dfrac{1,73}{1,25}}$}} };
\nodeshadowed [at={(14.5,4 )},yslant=0.0]
{ {\small \textcolor{black} {$\mathbf{\dfrac{6,33}{4,13}}$}} };
\nodeshadowed [at={(16.5,4 )},yslant=0.0]
{ {\small \textcolor{black} {$\mathbf{\dfrac{5,50}{3,18}}$}} };
\nodeshadowed [at={(14.5,2 )},yslant=0.0]
{ {\small \textcolor{black} {$\mathbf{\dfrac{9,27}{5,11}}$}} };

\draw[very thick,blue,->] (1.38,6.80) -- (1.82,6.80);
\draw[very thick,blue,->] (3.38,6.80) -- (3.82,6.80);
\draw[very thick,blue,->] (5.38,6.80) -- (5.82,6.80);
\draw[very thick,blue,->] (3.38,4.80) -- (3.82,4.80);
\draw[very thick,blue,->] (5.38,4.80) -- (5.82,4.80);
\draw[very thick,blue,->] (5.38,2.80) -- (5.82,2.80);
\draw[very thick,blue,->] (2.5,5.9) -- (2.5,5.46);
\draw[very thick,blue,->] (4.5,5.9) -- (4.5,5.46);
\draw[very thick,blue,->] (6.5,5.9) -- (6.5,5.46);
\draw[very thick,blue,->] (4.5,3.9) -- (4.5,3.46);
\draw[very thick,blue,->] (6.5,3.9) -- (6.5,3.46);
\draw[very thick,blue,->] (6.5,1.9) -- (6.5,1.46);

\draw[very thick,blue,->] (7.82,6.80) -- (7.38,6.80);
\draw[very thick,blue,->] (7.82,4.80) -- (7.38,4.80);
\draw[very thick,blue,->] (7.82,2.80) -- (7.38,2.80);
\draw[very thick,blue,->] (8.5,5.9) -- (8.5,5.46);
\draw[very thick,blue,->] (8.5,3.9) -- (8.5,3.46);

\draw[very thick,blue,->] (11.82,6.80) -- (11.38,6.80);
\draw[very thick,blue,->] (11.82,4.80) -- (11.38,4.80);
\draw[very thick,blue,->] (11.82,1.87) -- (11.38,1.44);
\draw[very thick,blue,->] (13.38,6.80) -- (13.82,6.80);
\draw[very thick,blue,->] (13.38,4.80) -- (13.82,4.80);
\draw[very thick,blue,->] (13.38,2.80) -- (13.82,2.80);
\draw[very thick,blue,->] (10.5,5.9) -- (10.5,5.46);
\draw[very thick,blue,->] (10.5,3.9) -- (10.5,3.46);
\draw[very thick,blue,->] (10.5,1.9) -- (10.5,1.46);

\draw[very thick,blue,->] (15.82,6.80) -- (15.38,6.80);
\draw[very thick,blue,->] (15.82,4.80) -- (15.38,4.80);
\draw[very thick,blue,->] (17.82,6.80) -- (17.38,6.80);
\draw[very thick,blue,->] (12.5,5.9) -- (12.5,5.46);
\draw[very thick,blue,->] (12.5,3.9) -- (12.5,3.46);
\draw[very thick,blue,->] (14.5,5.9) -- (14.5,5.46);
\draw[very thick,blue,->] (14.5,3.9) -- (14.5,3.46);
\draw[very thick,blue,->] (16.5,5.9) -- (16.5,5.46);

\draw[very thick, blue,->] (12.2,7.46) ..  controls +(up: .7cm)  and +(up: 1.2cm)  .. (8.8,7.46);
\draw[very thick, blue,->] (12.2,5.46) ..  controls +(up: .7cm)  and +(up: 1.2cm)  .. (8.8,5.46);
\draw[very thick, blue,->] (12.2,3.46) ..  controls +(up: .7cm)  and +(up: 1.2cm)  .. (8.8,3.46);

\end{tikzpicture}
\end{tabular}}
\end{center}
\end{table}
\end{center}
\newpage
\appendix\label{Appendix}
\section{The list of smooth $\IZ_3$ quotients of CICY's}
In this appendix, we take a second look at the three sequences that were given previously with the aim of presenting the changing structure of the CICY's that appear. Thus we start with the sequences that were previously given and then give further tables that present the diagrams for the CICY's of the sequence.

\def\str{\vrule height14pt depth8pt width0pt}
\setlength{\doublerulesep}{3pt}
\vspace{-10pt}
%
%
\begin{center}
\begin{table}[H]
\begin{center}
\capt{6.0in}{List1}{The first sub-web of $\IZ_3$ quotients generated by 
$\mathscr{X}^{1,16} = \mathscr{X}^{3,48}/ \IZ_3$.}
\addtocounter{table}{-1}
\vskip10pt
\framebox[6.2in]{
\begin{tabular}{c}
\begin{tikzpicture}[scale=1.2]
\clip (-2.5, -.4) rectangle (9.55,5.7);
\def\nodeshadowed[#1]#2;{\node[scale=1.1,above,#1]{#2};}

\nodeshadowed [at={(-1,0 )},yslant=0.0]
{ {\small \textcolor{black} {$\mathbf{\left( \mathscr{X}^{6,24}/ \IZ_3 \right) ^{4,10}}$}} };
\nodeshadowed [at={(2,0 )},yslant=0.0]
{ {\small \textcolor{black} {$\mathbf{\color{red} \left( \mathscr{X}^{9,21}/ \IZ_3 \right) ^{5,9}}$ }} };
\nodeshadowed [at={(5,0 )},yslant=0.0]
{ {\small \textcolor{black} {$\mathbf{\color{red} \left( \mathscr{X}^{12,18}/ \IZ_3 \right) ^{6,8}}$ }} };
\nodeshadowed [at={(8,0 )},yslant=0.0]
{ {\small \textcolor{black} {$\mathbf{\left( \mathscr{X}^{15,15}/ \IZ_3 \right) ^{7,7}}$}} };
\nodeshadowed [at={(-1,1.5 )},yslant=0.0]
{ {\small \textcolor{black} {$\mathbf{\left( \mathscr{X}^{5,32}/ \IZ_3 \right) ^{3,12}}$ }} };
\nodeshadowed [at={(2,1.5 )},yslant=0.0]
{ {\small \textcolor{black} {$\mathbf{\color{red}  \left( \mathscr{X}^{8,29}/ \IZ_3 \right) ^{4,11}}$}} };
\nodeshadowed [at={(5,1.5 )},yslant=0.0]
{ {\small \textcolor{black} {$\mathbf{\color{red} \left( \mathscr{X}^{11,26}/ \IZ_3 \right) ^{5,10}}$}} };
\nodeshadowed [at={(-1,3 )},yslant=0.0]
{ {\small \textcolor{black} {$\mathbf{\left( \mathscr{X}^{4,40}/ \IZ_3 \right) ^{2,14}}$}} };
\nodeshadowed [at={(2,3 )},yslant=0.0]
{ {\small \textcolor{black} {$\mathbf{\color{red} \left(  \mathscr{X}^{7,37} / \IZ_3 \right) ^{3,13}}$}} };
\nodeshadowed [at={(-1,4.5 )},yslant=0.0]
{ {\small \textcolor{black} {$\mathbf{\left( \mathscr{X}^{3,48}/ \IZ_3 \right) ^{1,16}}$}} };

\draw[very thick,blue,->] (-1, 4.5) -- (-1,3.7);
\draw[very thick,blue,->] (-1, 3.) -- (-1,2.2);
\draw[very thick,blue,->] (-1, 1.5) -- (-1,.7);
\draw[very thick,blue,->] (2, 3.) -- (2,2.2);
\draw[very thick,blue,->] (2, 1.5) -- (2,.7);
\draw[very thick,blue,->] (5, 1.5) -- (5,.7);

\draw[very thick,blue,->] (-.1, 3.3) -- (.8, 3.3);
\draw[very thick,blue,->] (-.1, 1.8) -- (.8, 1.8);
\draw[very thick,blue,->] (-.1, .3) -- (.8, .3);
\draw[very thick,blue,->] (2.9, 1.8) -- (3.74, 1.8);
\draw[very thick,blue,->] (2.9, .3) -- (3.74, .3);
\draw[very thick,blue,->] (5.9, .3) -- (6.8, .3);

\end{tikzpicture}
\end{tabular}}
\end{center}
\end{table}
\end{center}
\vspace{-10pt}
\begin{center}
\begin{longtable}{|c|c|c|}

\hline \multicolumn{1}{|c|}{\str\textbf{CICY Manifold}}&  \multicolumn{1}{|c|}{\textbf{$\IZ_3-$Quotient}} & \multicolumn{1}{|c|}{\textbf{CICY Diagram}} \\ \hline 
\endfirsthead


\hline\multicolumn{1}{|c|}{\str\textbf{CICY Manifold}} &
\multicolumn{1}{|c|}{\textbf{$\IZ_3-$Quotient}} &
\multicolumn{1}{|c|}{\textbf{CICY Diagram}} \\ \hline 
\endhead

\hline\hline \multicolumn{3}{|r|}{{\str Continued on next page}} \\ \hline
\endfoot

\hline\hline\multicolumn{3}{|c|}{\str}\\ \hline
\endlastfoot

\hline
\hline
\multicolumn{3}{|l|}{\str\textbf{First Row}}\\
\hline
\vrule height28pt  width0pt depth20pt  $\mathscr{X}^{3,48}$ & $\mathscr{X}^{1,16}$ & \lower16pt \hbox{\includegraphics[width=.85in]{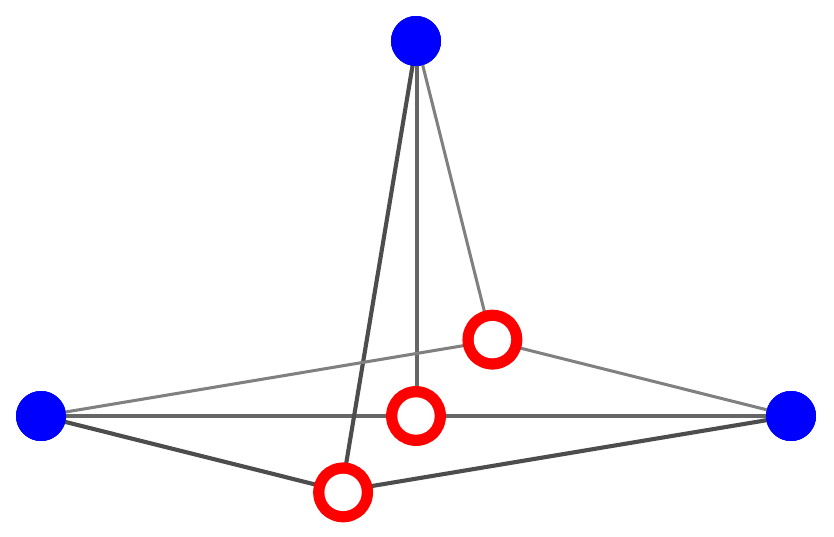}}  \\
\hline
\hline
\multicolumn{3}{|l|}{\str\textbf{Second Row}}\\
\hline
\vrule height30pt  width0pt depth20pt  $\mathscr{X}^{4,40}$ & $\mathscr{X}^{2,14}$ &\hskip .4in \lower16pt \hbox{\includegraphics[width=1.275in]{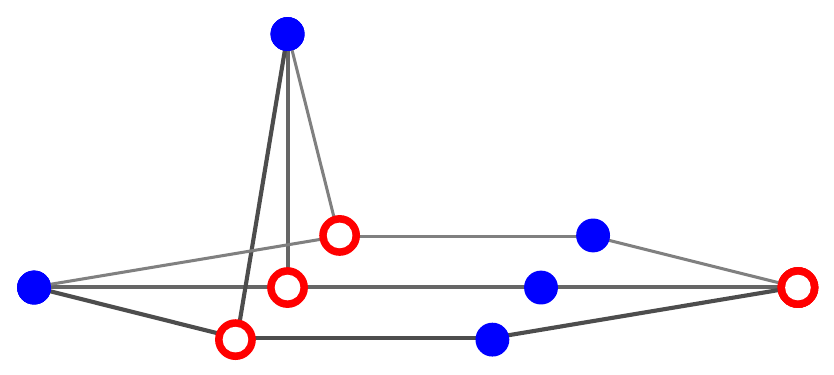}}  \\
\hline
\vrule height32pt  width0pt depth20pt  $\mathscr{X}^{7,37}$ & $\mathscr{X}^{3,13}$ &\hskip 1.2in \lower16pt \hbox{\includegraphics[width=2.125in]{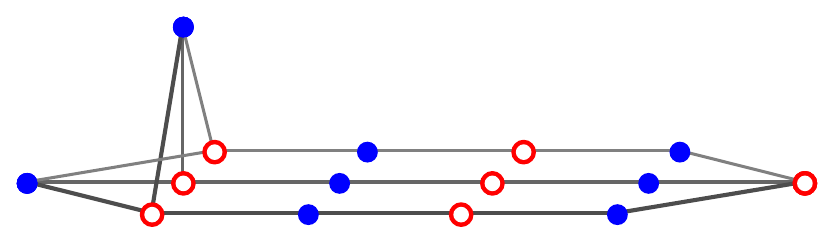}}  \\
\hline
\hline
\multicolumn{3}{|l|}{\str\textbf{Third Row}}\\
\hline
\vrule height30pt  width0pt depth20pt  $\mathscr{X}^{5,32}$ & $\mathscr{X}^{3,12}$ &\hskip 0in \lower16pt \hbox{\includegraphics[width=1.7in]{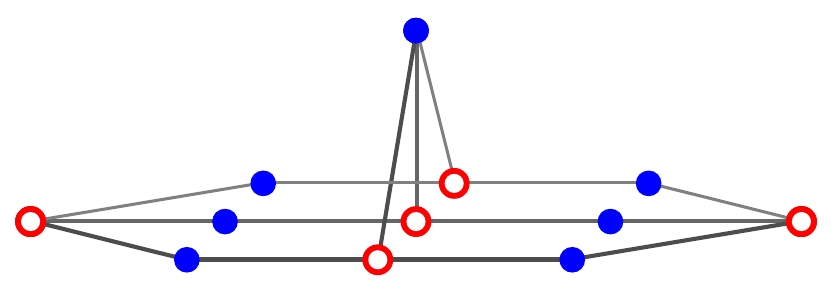}}  \\
\hline
\vrule height32pt  width0pt depth20pt  $\mathscr{X}^{8,29}$ & $\mathscr{X}^{4,11}$ &\hskip .8in \lower16pt \hbox{\includegraphics[width=2.55in]{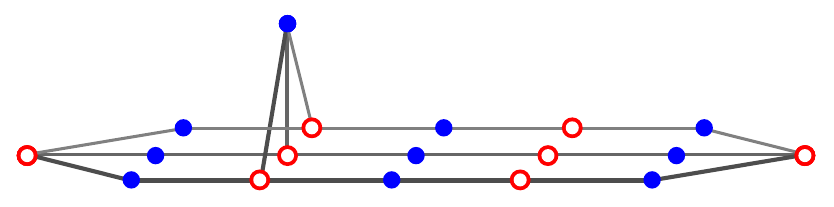}}  \\
\hline
\vrule height32pt  width0pt depth20pt  $\mathscr{X}^{11,26}$ & $\mathscr{X}^{5,10}$ &\hskip 0in \lower16pt \hbox{\includegraphics[width=3.4in]{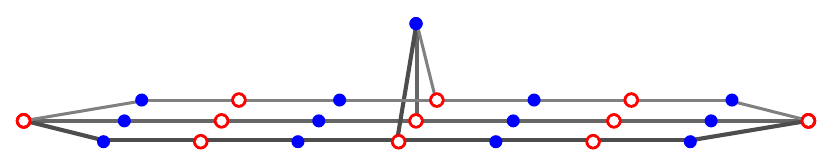}}  \\
\hline
\hline
\multicolumn{3}{|l|}{\str\textbf{Fourth Row}}\\
\hline
\vrule height61pt  width0pt depth20pt \lower -14pt \hbox{$\mathscr{X}^{6,24}$} & \lower -14pt \hbox{$\mathscr{X}^{4,10}$} &\hskip 0in \lower16pt \hbox{\includegraphics[width=1.7in]{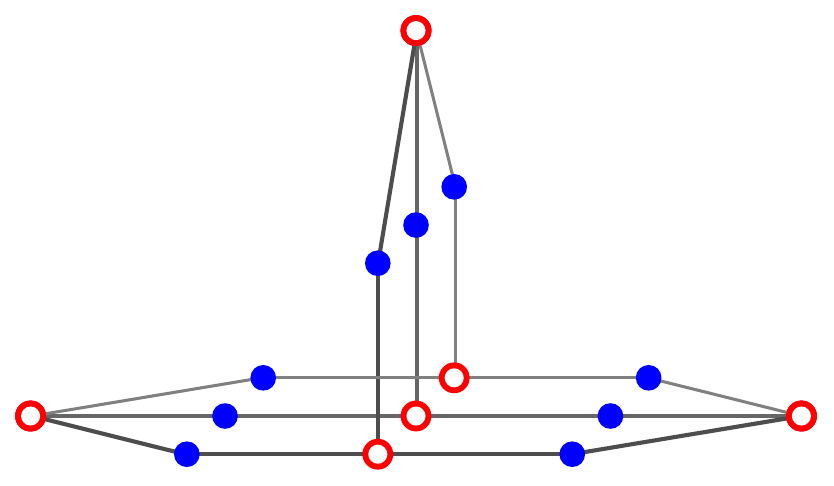}}  \\
\hline
\vrule height61pt  width0pt depth20pt  \lower -14pt \hbox{$\mathscr{X}^{9,21}$} & \lower -14pt \hbox{$\mathscr{X}^{5,9}$} &\hskip .8in \lower16pt \hbox{\includegraphics[width=2.55in]{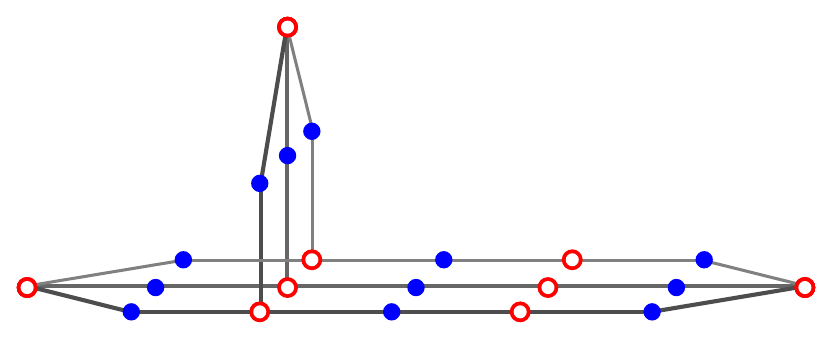}}  \\
\hline
\vrule height61pt  width0pt depth20pt  \lower -14pt  \hbox{$\mathscr{X}^{12,18}$} & \lower -14pt \hbox{$\mathscr{X}^{6,8}$} &\hskip 0in \lower16pt \hbox{\includegraphics[width=3.4in]{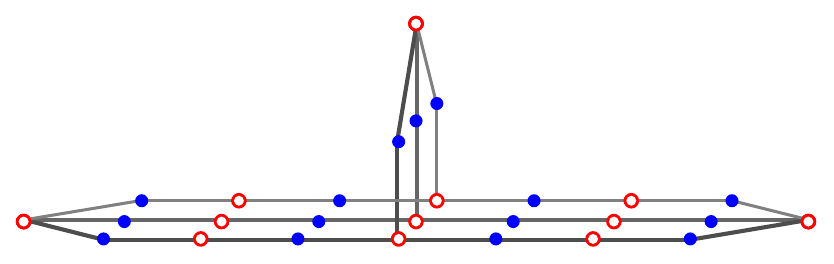}}  \\
\hline
\vrule height81pt  width0pt depth20pt  \lower -44pt \hbox{$\mathscr{X}^{15,15}$} & \lower -44pt \hbox{$\mathscr{X}^{7,7}$} &\hskip 0in \lower16pt \hbox{\includegraphics[width=3.4in]{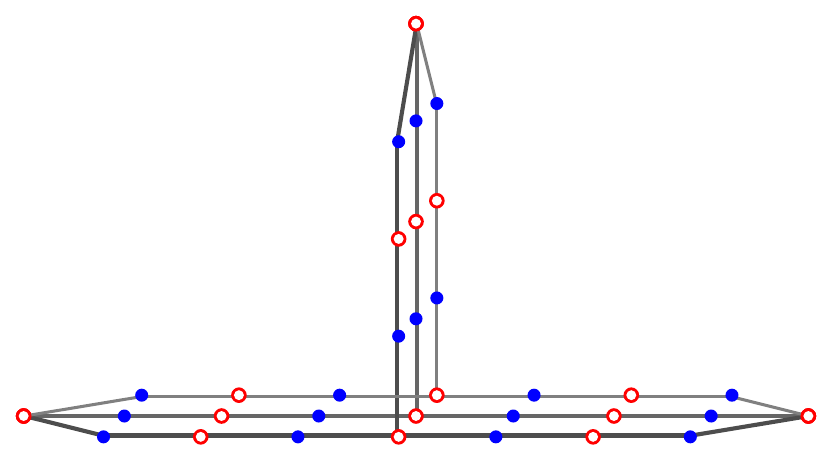}}  
\end{longtable}
\end{center}
%
%
\vspace{-60pt}
\begin{center}
\begin{table}[H]
\begin{center}
\capt{6.0in}{List3}{The second sub-web of $\IZ_3$ quotients generated by 
$\mathscr{X}^{1,25} = \mathscr{X}^{1,73}/ \IZ_3$.}
\addtocounter{table}{-1}
\vskip10pt
\framebox[6.2in]{
\begin{tabular}{c}
\begin{tikzpicture}[scale=1.2]
\clip (-1.3, 1.3) rectangle (8.35,5.5);
\def\nodeshadowed[#1]#2;{\node[scale=1.1,above,#1]{#2};}

\nodeshadowed [at={(.5,1.5 )},yslant=0.0]
{ {\small \textcolor{black} {$\mathbf{\left( \mathscr{X}^{3,39} / \IZ_3 \right) ^{ 3,15}} $}} };
\nodeshadowed [at={(3.5,1.5 )},yslant=0.0]
{ {\small \textcolor{black} {$\mathbf{\left( \mathscr{X}^{6,33} / \IZ_3 \right) ^{4,13}} $}} };
\nodeshadowed [at={(6.5,1.5 )},yslant=0.0]
{ {\small \textcolor{black} {$\mathbf{\left( \mathscr{X}^{9,27}/ \IZ_3 \right) ^{5,11}} $}} };
\nodeshadowed [at={(.5,3 )},yslant=0.0]
{ {\small \textcolor{black} {$\mathbf{\left( \mathscr{X}^{2,56} / \IZ_3 \right) ^{ 2,20}}$}} };
\nodeshadowed [at={(3.5,3 )},yslant=0.0]
{ {\small \textcolor{black} {$\mathbf{\left( \mathscr{X}^{5,50} / \IZ_3 \right) ^{ 3,18}} $}} };
\nodeshadowed [at={(.5,4.5 )},yslant=0.0]
{ {\small \textcolor{black} {$\mathbf{\left( \mathscr{X}^{1,73}/ \IZ_3 \right) ^{ 1,25}}$}} };

\draw[very thick,blue,->] (.5, 4.5) -- (.5,3.7);
\draw[very thick,blue,->] (.5, 3.) -- (.5,2.2);
\draw[very thick,blue,->] (3.5, 3.) -- (3.5,2.2);

\draw[very thick,blue,->] (1.4, 3.3) -- (2.3, 3.3);
\draw[very thick,blue,->] (1.4, 1.8) -- (2.3, 1.8);
\draw[very thick,blue,->] (4.4, 1.8) -- (5.24, 1.8);

\end{tikzpicture}
\end{tabular}}
\end{center}
\end{table}
\end{center}
\vspace{-15pt}
\begin{center}
\begin{longtable}{|c|c|c|}

\hline \multicolumn{1}{|c|}{\str\textbf{CICY Manifold}}&  \multicolumn{1}{|c|}{\textbf{$\IZ_3-$Quotient}} & \multicolumn{1}{|c|}{\textbf{CICY Diagram}} \\ \hline 
\endfirsthead


\hline \multicolumn{1}{|c|}{\str\textbf{CICY Manifold}} &
\multicolumn{1}{|c|}{\textbf{$\IZ_3-$Quotient}} &
\multicolumn{1}{|c|}{\textbf{CICY Diagram}} \\ \hline 
\endhead

\hline\hline \multicolumn{3}{|r|}{{\str Continued on next page}} \\ \hline
\endfoot

\hline\hline\multicolumn{3}{|c|}{\str}\\ \hline
\endlastfoot

\hline
\hline
\multicolumn{3}{|l|}{\str\textbf{First Row}}\\
\hline
\vrule height28pt  width0pt depth20pt  $\mathscr{X}^{1,73}$ & $\mathscr{X}^{1,25}$ & \lower4pt \hbox{\includegraphics[width=.85in]{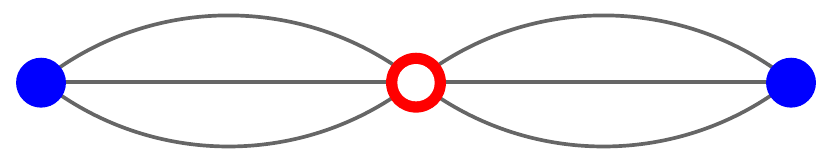}}  \\
\hline
\hline
\multicolumn{3}{|l|}{\str\textbf{Second Row}}\\
\hline
\vrule height32pt  width0pt depth20pt  $\mathscr{X}^{2,56}$ & $\mathscr{X}^{2,20}$ &\hskip .4in \lower10pt \hbox{\includegraphics[width=1.275in]{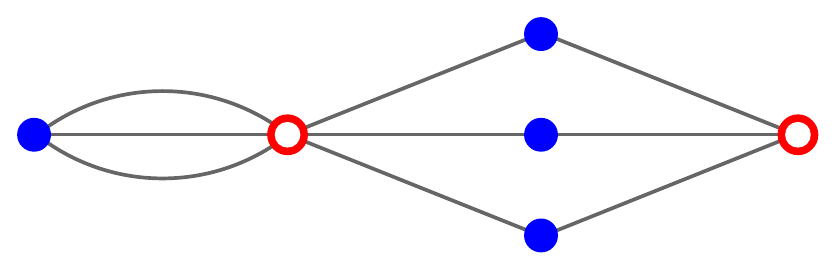}}  \\
\hline
\vrule height32pt  width0pt depth20pt  $\mathscr{X}^{5,50}$ & $\mathscr{X}^{3,18}$ &\hskip 1.20in \lower12pt \hbox{\includegraphics[width=2.125in]{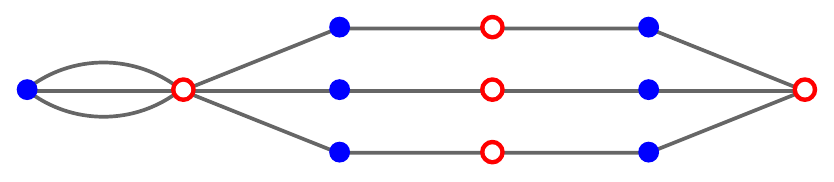}}  \\
\hline
\hline
\multicolumn{3}{|l|}{\str\textbf{Third Row}}\\
\hline
\vrule height30pt  width0pt depth20pt  $\mathscr{X}^{3,39}$ & $\mathscr{X}^{3,15}$ &\hskip .0in \lower12pt \hbox{\includegraphics[width=1.7in]{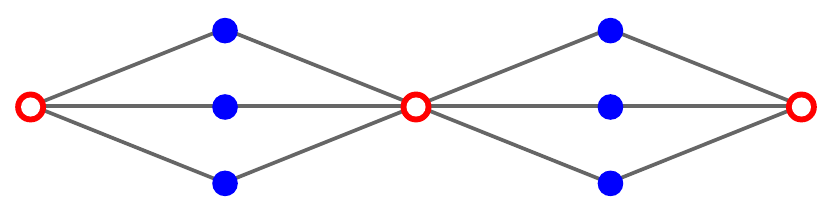}}  \\
\hline
\vrule height32pt  width0pt depth20pt  $\mathscr{X}^{6,33}$ & $\mathscr{X}^{4,13}$ &\hskip .8in \lower12pt \hbox{\includegraphics[width=2.55in]{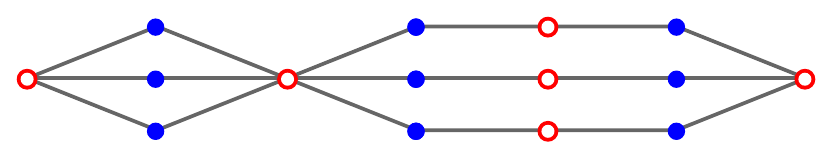}}  \\
\hline
\vrule height32pt  width0pt depth20pt  $\mathscr{X}^{9,27}$ & $\mathscr{X}^{5,11}$ &\hskip 0in \lower12pt \hbox{\includegraphics[width=3.4in]{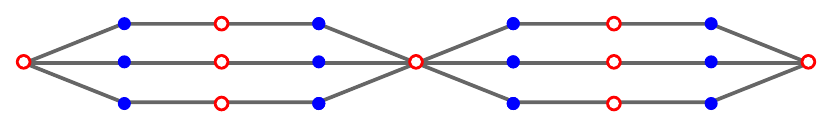}}  \\
\end{longtable}
\end{center}

%
%
\vspace{-60pt}
\begin{center}
\begin{table}[H]
\begin{center}
\capt{6.0in}{List2}{The third sub-web of $\IZ_3$ quotients generated by 
$\mathscr{X}^{2,29} = \mathscr{X}^{2,83}/ \IZ_3$.}
\vskip10pt
\framebox[6.2in]{
\begin{tabular}{c}
\begin{tikzpicture}[scale=1.2]
\clip (-2.5, -.2) rectangle (9.55,4.3);
\def\nodeshadowed[#1]#2;{\node[scale=1.1,above,#1]{#2};}

\nodeshadowed [at={(-1,3 )},yslant=0.0]
{ {\small \textcolor{black} {$\mathbf{\left( \mathscr{X}^{3,48}/\IZ_3 \right)^{3,18}}$}} };
\nodeshadowed [at={(2,3)},yslant=0.0]
{ {\small \textcolor{black} {$\mathbf{\left( \mathscr{X}^{6,33} /\IZ_3 \right)^{4,13}}$}} };
\nodeshadowed [at={(5,3)},yslant=0.0]
{ {\small \textcolor{black} {$\mathbf{\left( \mathscr{X}^{9,21} /\IZ_3 \right)^{5,9}}$}} };

\nodeshadowed [at={(-1,1.5 )},yslant=0.0]
{ {\small \textcolor{black} {$\mathbf{\left( \mathscr{X}^{2,83} /\IZ_3 \right)^{2,29}}$}} };
\nodeshadowed [at={(2,1.5 )},yslant=0.0]
{ {\small \textcolor{black} { $\mathbf{\left( \mathscr{X}^{5,59} /\IZ_3 \right)^{3,21}}$}} };
\nodeshadowed [at={(5,1.5 )},yslant=0.0]
{ {\small \textcolor{black} {$\mathbf{\left( \mathscr{X}^{8,44} /\IZ_3 \right)^{4,16}}$}} };

\nodeshadowed [at={(-1,0 )},yslant=0.0]
{ {\small \textcolor{black} {$\mathbf{\left( \mathscr{X}^{8,35}/\IZ_3 \right)^{4,13} }$}} };
\nodeshadowed [at={(2,0 )},yslant=0.0]
{ {\small \textcolor{black} {$\mathbf{\color{red} \left( \mathscr{X}^{11,29}/\IZ_3 \right)^{ 5,11} }$}} };
\nodeshadowed [at={(5,0 )},yslant=0.0]
{ {\small \textcolor{black} {$\mathbf{\left( \mathscr{X}^{14,23}/\IZ_3 \right)^{ 6,9}}$ }} };
\nodeshadowed [at={(8,0 )},yslant=0.0]
{ {\small \textcolor{black} {$\mathbf{\left( \mathscr{X}^{19,19} /\IZ_3 \right)^{7,7} }$}} };

\draw[very thick,blue,->] (-.1, 3.3) -- (.74, 3.3);
\draw[very thick,blue,->] (2.9, 3.3) -- (3.74, 3.3);
\draw[very thick,blue,->] (-.1, 1.8) -- (.74, 1.8);
\draw[very thick,blue,->] (2.9, 1.8) -- (3.74, 1.8);
\draw[very thick,blue,->] (-.1, .3) -- (.74, .3);
\draw[very thick,blue,->] (2.9, .3) -- (3.74, .3);
\draw[very thick,blue,->] (5.9, .3) -- (6.77, .3);

\draw[very thick,blue,->] (-1, 2.25) -- (-1, 2.95);
\draw[very thick,blue,->] (2, 2.25) -- (2, 2.95);
\draw[very thick,blue,->] (5, 2.25) -- (5, 2.95);

\draw[very thick,blue,->] (-1, 1.45) -- (-1, .75);
\draw[very thick,blue,->] (2, 1.45) -- (2, .75);
\draw[very thick,blue,->] (5.9, 1.45) -- (6.8, .75);
\end{tikzpicture}
\end{tabular}}
\end{center}
\end{table}
\end{center}
\begin{center}
\begin{longtable}{|c|c|c|}

\hline \multicolumn{1}{|c|}{\str\textbf{CICY Manifold}}&  \multicolumn{1}{|c|}{\textbf{$\IZ_3-$Quotient}} & \multicolumn{1}{|c|}{\textbf{CICY Diagram}} \\ \hline 
\endfirsthead
%
%
\hline \multicolumn{1}{|c|}{\str\textbf{CICY Manifold}} &
\multicolumn{1}{|c|}{\textbf{$\IZ_3-$Quotient}} &
\multicolumn{1}{|c|}{\textbf{CICY Diagram}} \\ \hline 
\endhead
\hline\hline \multicolumn{3}{|r|}{{\str Continued on next page}} \\ \hline
\endfoot
\hline\hline\multicolumn{3}{|c|}{\str}\\ \hline
\endlastfoot
\hline\hline
\multicolumn{3}{|l|}{\str\textbf{First Row}}\\
\hline
\vrule height28pt  width0pt depth20pt  $\mathscr{X}^{3,48}$ & $\mathscr{X}^{3,18}$ & \lower16pt \hbox{\includegraphics[width=.85in]{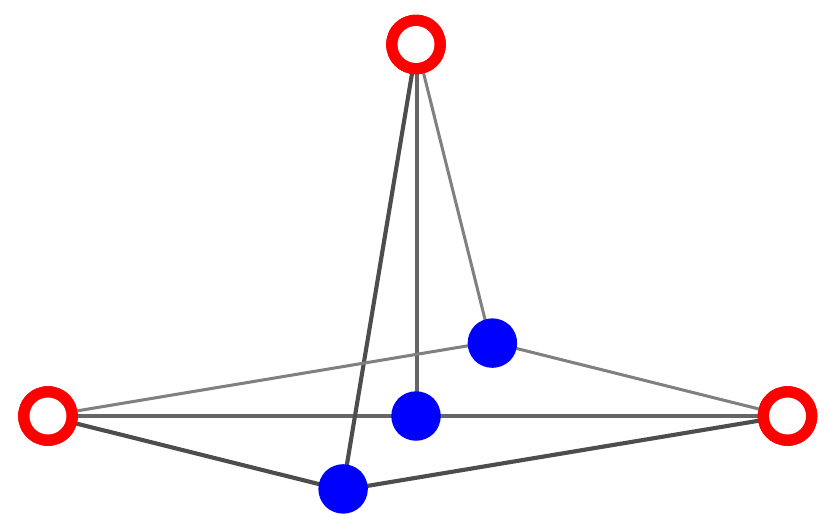}}  \\
\hline
\vrule height32pt  width0pt depth20pt  $\mathscr{X}^{6,33}$ & $\mathscr{X}^{4,13}$ &\hskip .8in \lower16pt \hbox{\includegraphics[width=1.7in]{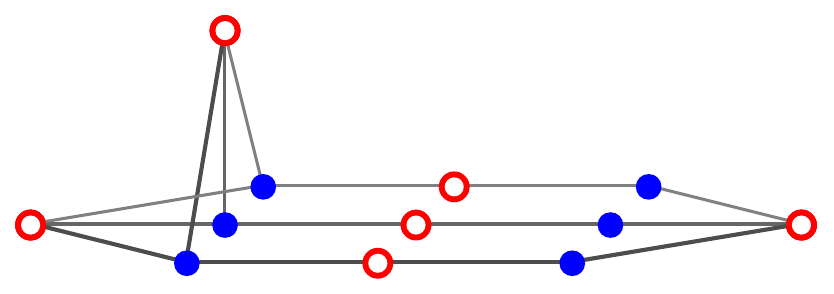}}  \\
\hline
\vrule height32pt  width0pt depth20pt  $\mathscr{X}^{9,21}$ & $\mathscr{X}^{5,9}$ &\hskip 0in \lower16pt \hbox{\includegraphics[width=2.55in]{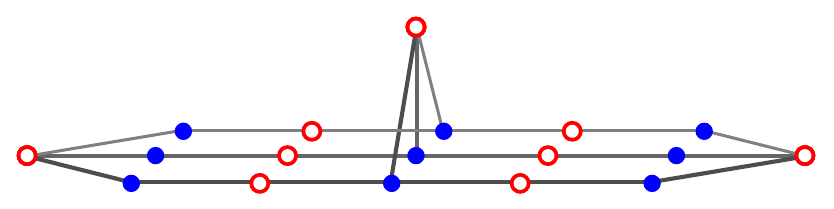}}  \\
\hline
\hline
\multicolumn{3}{|l|}{\str\textbf{Second Row}}\\
\hline
\vrule height30pt  width0pt depth20pt  $\mathscr{X}^{2,83}$ & $\mathscr{X}^{2,29}$ &\hskip 0in \lower 4pt \hbox{\includegraphics[width=.85in]{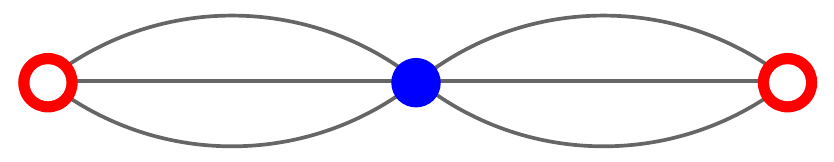}}  \\
\hline
\vrule height32pt  width0pt depth20pt  $\mathscr{X}^{5,59}$ & $\mathscr{X}^{3,21}$ &\hskip .8in \lower10pt \hbox{\includegraphics[width=1.7in]{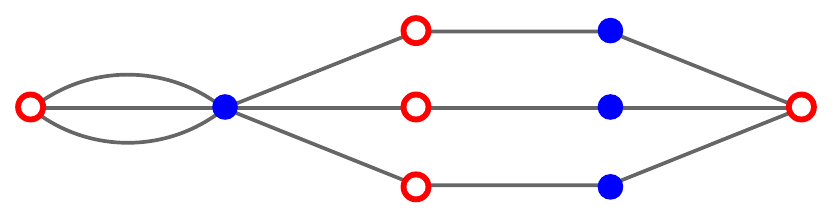}}  \\
\hline
\vrule height32pt  width0pt depth20pt  $\mathscr{X}^{8,44}$ & $\mathscr{X}^{4,16}$ &\hskip 0in \lower12pt \hbox{\includegraphics[width=2.55in]{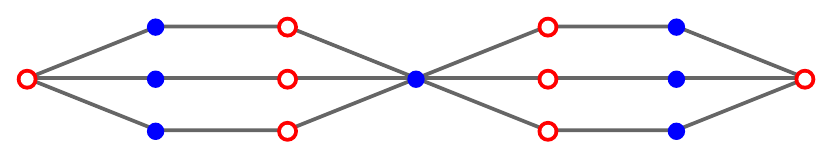}}  \\
\hline
\hline
\multicolumn{3}{|l|}{\str\textbf{Third Row}}\\
\hline
\vrule height30pt  width0pt depth20pt  $\mathscr{X}^{8,35}$ & $\mathscr{X}^{4,13}$ &\hskip 0in \lower12pt \hbox{\includegraphics[width=1.7in]{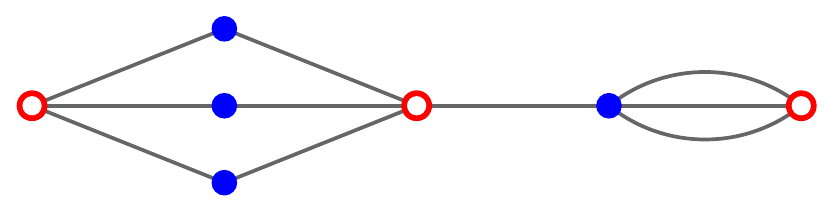}}  \\
\hline
\vrule height32pt  width0pt depth20pt  $\mathscr{X}^{11,29}$ & $\mathscr{X}^{5,11}$ &\hskip .8in \lower16pt \hbox{\includegraphics[width=2.55in]{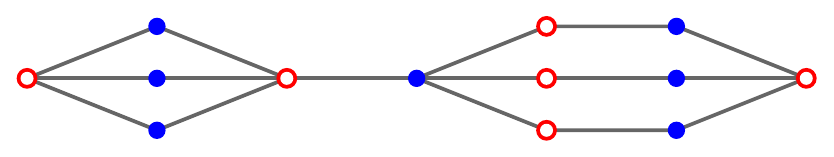}}  \\
\hline
\vrule height32pt  width0pt depth20pt  $\mathscr{X}^{14,23}$ & $\mathscr{X}^{6,9}$ &\hskip .8in \lower16pt \hbox{\includegraphics[width=2.5in]{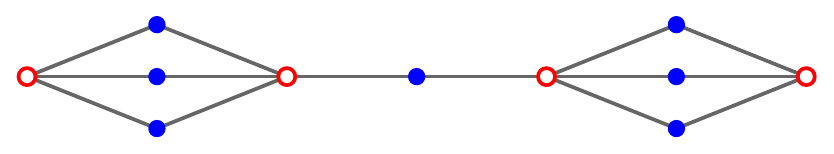}}  \\
\hline
\vrule height32pt  width0pt depth20pt  $\mathscr{X}^{19,19}$ & $\mathscr{X}^{7,7}$ &\hskip 0in \lower16pt \hbox{\includegraphics[width=3.4in]{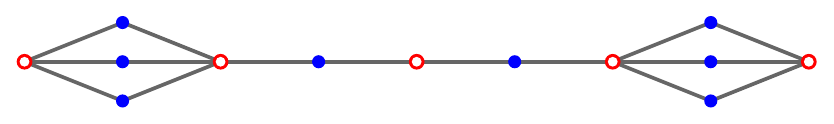}}  \\
\end{longtable}
\end{center}
%
%
\vskip20pt
\section*{Acknowledgements}
We are indebted to Rhys Davies for many useful conversations, and particularly for his comments on conifold transitions between free quotients of Calabi-Yau threefolds. Volker Braun was kind enough to comment on the manuscript. Andr\'e~Lukas and James Gray provided assistance with the installation and usage of the STRINGVACUA package. AC is supported in part by a Graduate Scholarship from University College in association with the Rudolf Peierls Centre for Theoretical Physics, University of Oxford. 

\newpage

\end{document}